\documentstyle[amsfonts,epsf,10pt]{article}
\input{epsf}
\renewcommand{\theequation}{\arabic{section}.\arabic{equation}}

\newcommand{\reell}{\kern+.23em\sf{I}\kern-.40em\sf{R}\kern+.76em\kern-.25em}

\def\n3{n_{3}}

\newcommand{\be}{\begin{equation}}
\newcommand{\ee}{\end{equation}}
\newcommand{\ba}{\begin{eqnarray}}
\newcommand{\ea}{\end{eqnarray}}

\begin{document}

\title{Quantization of Multidimensional Cat Maps}
\author{{A.M.F. Rivas$^1$\thanks{%
Corresponding author. e-mail:rivas@cbpf.br, Tel:(5521)5867223
fax:(5521)5867400 }, M. Saraceno$^2$ and A.M. Ozorio de Almeida$^1$}}
\maketitle

\noindent

\centerline{\it $^1$Centro Brasileiro de Pesquisas F\'{\i}sicas} \centerline%
{\it Rua Xavier Sigaud 150, CEP 22290-180, RJ, Rio de Janeiro, Brazil} %
\centerline{\it $^2$Comisi\'{o}n Nacional de Energ\'{\i}a At\'{o}mica
(CNEA), }\centerline{\it Ave. del Libertador 8250, 1429 Buenos Aires,
Argentina.}

\vspace{1cm} \noindent

PACS:


\centerline {\bf Abstract} 

In this work we study cat maps with many degrees of freedom. Classical cat maps are classified using  the 
Cayley parametrization of symplectic matrices and the closely associated
center and chord generating functions. Particular attention is dedicated to  loxodromic behavior, which is a new feature
of two-dimensional maps. The maps are then quantized using a recently
developed Weyl representation on the torus and the general condition on the
Floquet angles is derived for a particular map to be quantizable. The
semiclassical approximation is exact, regardless of the dimensionality or
of the nature of the fixed points.

\vfill
\clearpage

\section{Introduction}

Linearization of a dynamical system near a periodic orbit is one of the most
fruitful starting points for the analysis of classical motion. In its turn,
the symplectic group of linear Hamiltonian systems in plane phase space is
easily quantized to form the corresponding metaplectic quantum group.
Essentially, the generating function for the group of canonical
transformations is simply exponentiated to obtain a representation of the
quantum unitary transformation.

If the chosen orbit is a point of equilibrium, the corresponding linear
system belongs to the homogeneous symplectic group, characterized by a
single equilibrium, usually taken as the origin. Likewise, the Poincar\'{e}
map in the neighborhood of a periodic orbit is linearized into a homogeneous
symplectic map with discrete time. The essential character of the motion is
classified according to the eigenvalues of the symplectic matrix ${\cal M}$, that
determines the evolution of phase space points $x:$%
\begin{equation}
x^{\prime }={\cal M}x.
\end{equation}
There may be

\begin{description}
\item[(a)]  pairs of eigenvalues ($\lambda ,\lambda ^{-1});$

\item[(b)]  pairs of eigenvalues on the unit circle ($e^{i\theta
},e^{-i\theta })$

\item[(c)]  quartets of general complex eigenvalues $\lambda ^{\pm 1}e^{\pm
i\theta }.$
\end{description}

\noindent On varying parameters, it is possible to obtain unit eigenvalues ,
or eigenvalue collisions, but the above classification is generic for a
given symplectic system \cite{quinze}.

It is always possible to decompose such a generic linearized system into
sub-systems in invariant subspaces of two dimensions, corresponding to cases
(a)\ and (b) above, or four dimensions in case (c). Case (b) is the {\it %
elliptic map, }which is trivially integrable, whereas case (a)\ defines {\it %
hyperbolic motion.} This is also very simple in the linear limit, but can
become a source of chaotic mixing as nonlinear perturbations are added.
Alternatively, this effect is achieved by wrapping the plane space itself
into a torus.

The resulting symplectomorphism of the torus is known colloquially as a {\it %
cat map}, characterized by a symplectic matrix with integer elements. A
hyperbolic cat map is structurally stable, i.e. the orbit structure is
invariant with respect to small nonlinear perturbations as a consequence of
Anosov's theorem \cite{quinze}. The same is true of a four dimensional {\it %
loxodromic cat map} with general complex eigenvalues in case (c). These
structurally invariant systems are known as {\it Anosov} systems, they are
ergodic and mixing.

It follows that four dimensions is the lower bound in which we can study
loxodromic periodic orbits \cite{quinze}, characterized by stable and
unstable manifolds where the orbits spiral inwards and outwards
respectively, and their effect on the quantum energy spectrum. This is the
reason for their absence in all previous studies of the quantization of cat
maps, though Greenman \cite{greenman} has recently analyzed the periodic
structure of higher dimensional classical cat maps. Dimension four is also
the least dimension for the analysis of the decomposition of the
neighborhood of orbits into elliptic and (real)\ hyperbolic components.

In some cases this decomposition is only local, because the canonical
transformation that achieves it is not itself a cat map. Then the quantum
quasienergy spectrum will not be decomposed into the corresponding lower
dimensional spectra. In any case, all cat maps derived from each other as a
result of similarity transformations involving other cat maps are
equivalent: they have the same (classical and quantum) eigenvalues and the
same number of fixed points. ( Note that, on the torus, a homogeneous linear
map may have multiple fixed points.)

For this reason, we discuss in the next section the integer subgroup of
symplectic transformations, nicknamed the feline group. For higher
dimensions than two, we encounter the problem of a priori identification of
a cat map. An alternative approach involves generating functions, of which
there are several choices. However there is a great advantage to using
generating functions that are invariant with respect to feline
transformations.

In section 3 we analyze the dynamics of classical cat maps and classify
four-dimensional cat maps, providing examples of various types. These
examples are then quantized in section 4. They share a simplifying property
that permits us to discuss the periodicity of the propagator and the
exactness of the Gutzwiller trace formula, without analyzing the subtleties
of general cat map quantization.  This is the subject of section 5, where we determine
the set of Floquet angles that allow the quantization of a given map; and study the feline invariance of the quantization. 

\section{ Feline-invariant generating functions}

\setcounter{equation}{0}

A point in the even-dimensional {\it phase space} with $L$ degrees of
freedom on a $2L$-torus has coordinates separated into $L$ momenta and $L$
positions, so that $x=\left( 
\begin{array}{c}
p \\ 
q
\end{array}
\right) =\left( 
\begin{array}{c}
p_1,\cdots ,p_L \\ 
q_1,\cdots ,q_L
\end{array}
\right) $. All the $2L$ coordinates are periodic with periods $\Delta q_i$
and $\Delta p_i$. For simplicity we will treat the case where we can choose
units so that $\Delta q_i$ and $\Delta p_i$ are all equal to 1. The range of
values of $x$ is then the unit $2L$-hypercube denoted from now on by $\Box .$

Let us consider then a linear automorphism on the $2L$-torus generated by
the $2L\times 2L$ matrix ${\cal M}$, that takes a point $x_{-}=\left( 
\begin{array}{c}
p_{-} \\ 
q_{-}
\end{array}
\right) $ to a point $x_{+}=\left( 
\begin{array}{c}
p_{+} \\ 
q_{+}
\end{array}
\right) :$%
\begin{equation}
x_{+}={\cal M}x_{-}\quad \mbox{mod(1)}.  \label{mapa0}
\end{equation}
In other words, there exists an integer $2L$-dimensional vector ${\bf m}%
=\left( 
\begin{array}{l}
m_p \\ 
m_q
\end{array}
\right) ,$ such that 
\begin{equation}
x_{+}={\cal M}x_{-}-{\bf m.}  \label{mapa}
\end{equation}
The components of ${\bf m}$ denote the winding numbers made by the point $%
x_{-}$ around the respective irreducible circuit on the $2L$-torus after the
application of the map ${\cal M}.$ The torus will be divided into regions labeled
by their respective vector ${\bf m}$. For the map to be conservative, the ${\cal M}$
matrix must be symplectic, that is 
\begin{equation}
{\cal M}^t{\frak J}{\cal M}={\frak J,}  \label{msimple}
\end{equation}
where ${\cal M}^t$ is the transpose of ${\cal M}$ and 
\begin{equation}
{\frak J}=\left[ 
\begin{array}{c|c}
0 & -1 \\ \hline
1 & 0
\end{array}
\right] .
\end{equation}
The matrix ${\cal M}$ must have integer coefficients for the $2L$-torus to be
mapped onto itself.

For one degree of freedom ( $L=1$ ) these systems are known as Arnold's cat
maps \cite{quinze}. If $|tr({\cal M})|>2$ the map has two distinct real
eigenvectors, so it is ergodic, mixing and purely hyperbolic. For the case $%
|tr({\cal M})|<2$ there are no real eigenvectors, the map is then elliptic. For $%
|tr({\cal M})|=2$ there is only one real eigenvector and so the map is parabolic.
For more degrees of freedom richer structure may appear, as will be
discussed.

The set of matrices ${\cal M}$ satisfying (\ref{msimple}) form the symplectic
group, so that the matrix ${\cal M}^{\prime },$ obtained from ${\cal M}$ by the similarity
transformation 
\begin{equation}
{\cal N}^{-1}{\cal M}{\cal N}={\cal M}^{\prime }  \label{2.5}
\end{equation}
is also symplectic if the matrix ${\cal N}$ is itself symplectic and shares the
same eigenvalues as ${\cal M}$. Indeed, we may consider the symplectic
transformation corresponding to ${\cal M}^{\prime }$ as the same as ${\cal M}$, but viewed
in an alternative symplectic coordinate system.

Consider now a product of cat maps; this must be symplectic and all the
matrix elements will be integers. Since the inverse of a cat map is also a
symplectic integer transformation and the unit matrix likewise, the set of
cat maps is a subgroup of the symplectic group, appropriately nicknamed the 
{\it feline group}. Again, we may consider similarity transformations of the
form (\ref{2.5}) among cat maps ${\cal M}$ and ${\cal N}$ as defining essentially maps
viewed by alternative symplectic coordinates on the torus.

The importance of defining equivalence classes of similar cat maps increases
with the dimension of the torus. It is easy to define integer matrices and the {\it a posteriori} condition  (\ref{msimple}) merely restricts the value of
the determinant to unity when $L=1.$\ However, for $L=2$, the symplectic
property already implies ten independent conditions to be satisfied by the
sixteen integer matrix elements. An alternative procedure, that we adopt here, is to define the transformation
implicitly by means of a generating function, thus guaranteeing (\ref{msimple}). The 
disadvantage is now that we need to ensure that the resulting matrix ${\cal M}$ has
integer entries, leading to additional restrictions on the allowed generating
functions (see below ) . Nevertheless, this approach has proved very fruitful, because the quantization relies heavily on the generating function, and not on the transformation matrix itself.

The generating function in position representation for the one degree of
freedom case is \cite{keat1} 
\begin{equation}
S(q_{-},q_{+},{\bf m})=\frac 1{2{\cal M}_{21}}\left[
{\cal M}_{22}q_{-}^2-2q_{-}(q_{+}+m_q)+{\cal M}_{11}(q_{+}+m_q)^2-2{\cal M}_{21}m_pq_{+}\right] .
\end{equation}
This function generates the dynamics through

\begin{eqnarray}
p_{+} &=&\frac{\partial S(q_{-},q_{+},{\bf m})}{\partial q_{+}} \\
p_{-} &=&-\frac{\partial S(q_{-},q_{+},{\bf m})}{\partial q_{-}}.
\end{eqnarray}
Thus the quadratic part of the generating function is common to the entire
torus, whereas the linear part depends on the winding vector ${\bf m}$ that
changes discontinuously on the boundary of each subregion of the torus.

The weakness of the position generating function is that it transforms in a
complicated way under feline similarity transformations. Only in the special
case of a point transformation, that does not mix momenta with positions,
will the generating function remain invariant. In contrast, the {\it center
and chord generating functions}, are known to be symplectically invariant 
\cite{ozrep}. Adapted to the torus, we shall show that they can also be
chosen to be invariant under feline transformations.

The starting point is to define the center point $x$ and the chord $\xi $
such that 
\begin{equation}
x_{\pm }=x\pm \frac 12\xi ,  \label{qqxcor}
\end{equation}
so that the center 
\begin{equation}
x\equiv \frac{x_{+}+x_{-}}2  \label{xdef}
\end{equation}
and the chord 
\begin{equation}
\xi \equiv x_{+}-x_{-}.  \label{cordef}
\end{equation}
Given the initial point $x_{-}$, 
\begin{eqnarray}
x &=&\frac 12\left( {\cal M}+1\right) x_{-}-\frac{{\bf m}}2  \label{xx-} \\
\xi &=&\left( {\cal M}-1\right) x_{-}-{\bf m.}  \label{corx-}
\end{eqnarray}
Elimination of $x_{-}$, then establishes the direct relation between centers
and chords: 
\begin{eqnarray}
x &=&\frac 12\frac{\left( {\cal M}+1\right) }{({\cal M}-1)}\xi +({\cal M}-1)^{-1}{\bf m}
\label{xcor} \\
\xi &=&2\frac{\left( {\cal M}-1\right) }{\left( {\cal M}+1\right) }x-2\left( {\cal M}+1\right)
^{-1}{\bf m}.  \label{corx}
\end{eqnarray}

Center and chord generating functions respectively denoted by $S(x,{\bf m})$
and $S(\xi ,{\bf m})$ are defined in \cite{ozrep} so that the transformation
is obtained as 
\begin{eqnarray}
x &=&{\frak J}\frac{\partial S(\xi ,{\bf m})}{\partial \xi }  \label{xgenfu}
\\
\xi &=&-{\frak J}\frac{\partial S(x,{\bf m})}{\partial x}.  \label{corgenfu}
\end{eqnarray}
In analogy to the dynamics in the plane, we may consider that use of the
center representation identifies the orbit with the reflection (or
inversion) through $x$, because of (\ref{xdef}). Hence we shall also refer
to $x$ as the reflection point. The chord representation is locally
equivalent to the uniform translation of phase space by the chord (\ref
{cordef}). Equating (\ref{xcor}) with (\ref{xgenfu}) and (\ref{corx}) with (%
\ref{corgenfu}), we obtain the quadratic generating functions 
\begin{eqnarray}
S(x,{\bf m}) &=&xBx+x(B-{\frak J)}{\bf m}+f({\bf m})  \label{sxf} \\
S(\xi ,{\bf m}) &=&\frac 14\xi \beta \xi +\frac 12\xi (\beta +{\frak J)}{\bf %
m}+g({\bf m})  \label{scorg}
\end{eqnarray}
where $B$ and $\beta $ are symmetric matrices; they are the Cayley
parametrization of ${\cal M}:$%
\begin{eqnarray}
{\frak J}B &=&\frac{\left( 1-{\cal M}\right) }{\left( 1+{\cal M}\right) }  \label{jbm} \\
{\frak J}\beta &=&\frac{\left( {\cal M}+1\right) }{\left( {\cal M}-1\right) }
\label{jbetam}
\end{eqnarray}

If ${\cal M}$ has an eigenvalue equal to $1$ then the $\beta $ matrix will be
singular. This corresponds to a caustic of the center generating function.
Whereas if ${\cal M}$ has an eigenvalue equal to $-1$, then $B$ will be a singular
matrix which corresponds to a caustic of the chord generating function \cite
{ozrep}.Some useful relations obtained from (\ref{jbm}) and (\ref{jbetam})
are 
\begin{eqnarray}
(B-{\frak J}) &=&-2{\frak J}\left( 1+{\cal M}\right) ^{-1}, \\
(\beta +{\frak J}) &=&-2{\frak J}\left( {\cal M}-1\right) ^{-1},
\end{eqnarray}

\begin{equation}
{\frak J}B=-\frac 1{{\frak J}\beta }  \label{jbbeta}
\end{equation}
and 
\begin{equation}
{\cal M}=\frac{\left( 1-{\frak J}B\right) }{\left( 1+{\frak J}B\right) }=\frac{%
\left( {\frak J}\beta +1\right) }{\left( {\frak J}\beta -1\right) }.
\label{mbbeta}
\end{equation}

The functions $f({\bf m})$ and $g({\bf m})$ are arbitrary, since they only
depend on the winding number ${\bf m}$, so they do not affect the
transformation (\ref{xcor}) or (\ref{corx}). However, the center generating
function for cat maps can also be obtained directly from the map (\ref{mapa}%
). This is a composition of the symplectic map on the plane ${\cal M}$ , whose
center generating function is $S_1(x)=xBx,$ with the uniform translation $%
T_{-{\bf m}}$ of vector $-{\bf m}$ that pulls back the final point to the
unit cell $\Box .$ The generating function of such a translation is \cite
{ozrep} $S_2(x)=-{\bf m\wedge }x$ , also symplectically invariant. Then,
using the composition law for center generating functions \cite{ozrep} we
obtain the generating function (\ref{sxf}) with 
\begin{equation}
f({\bf m})=\frac 14{\bf m}B{\bf m.}  \label{3.30}
\end{equation}

As in the plane case \cite{ozrep} generating functions are related among
themselves by Legendre transformations. Thus, $S(x,{\bf m})$ is obtained
from the more familiar position generating function $S(q_{-},q_{+},{\bf m})$
as 
\begin{equation}
S(x,{\bf m})=S(q_{-},q_{+},{\bf m})+\frac 12(p_{-}+p_{+})(q_{+}-q_{-}),
\label{legqx}
\end{equation}
whereas the relation between the chord and center generating function is 
\begin{equation}
S(\xi ,{\bf m})=\xi \wedge x-S(x,{\bf m}).  \label{legxcor}
\end{equation}
In each case, the variable absent on the left side is eliminated by
requiring the right side to be stationary with respect to it. The skew
product in (\ref{legxcor}) , 
\begin{equation}
\xi \wedge \eta \ \equiv \sum_{\ell =1}^L\left( \xi _{p_\ell }\eta _{q_\ell
}-\xi _{q_\ell }\eta _{p_\ell }\right) =({\frak J}\xi ).\eta =-\xi {\frak J}%
\eta ,
\end{equation}
is the symplectic area of the parallelogram formed by any pair of vectors $%
\xi $ and $\eta $. Then, using (\ref{legxcor}), for the center function with
the term (\ref{3.30}), we obtain the chord generating function (\ref{scorg})
with 
\begin{equation}
g({\bf m})=\frac 14{\bf m}\beta {\bf m.}  \label{3.30g}
\end{equation}

We shall label periodic points of period $l$ as $x_l.$ Thus, fixed points $%
x_1$ are such that the chord $\xi =0,$ or the center $x=x_1=({\cal M}-1)^{-1}{\bf m,%
}$ which inserted in (\ref{legxcor}) leads to 
\begin{equation}
S(\xi =0,{\bf m})=-S(x_1,{\bf m}).  \label{spcorx}
\end{equation}
There follows the restriction that the terms in $S(\xi ,{\bf m})$ and $S(x,%
{\bf m})$ that depend only on ${\bf m}$ satisfy 
\begin{equation}
f({\bf m})+g({\bf m})=\frac 14{\bf m}(\beta +B){\bf m.}
\end{equation}
The choice (\ref{3.30}) and (\ref{3.30g}) obviously satisfy this criterion,
but another possibility is 
\begin{equation}
f({\bf m})=\frac 14{\bf m}(B+\widetilde{{\frak J}}){\bf m\quad }\mbox{and}%
\quad g({\bf m})=\frac 14{\bf m}(\beta -\widetilde{{\frak J}}){\bf m,}
\label{gm}
\end{equation}
where we define the symmetric matrix 
\begin{equation}
\widetilde{{\frak J}}=\left[ 
\begin{array}{c|c}
0 & 1 \\ \hline
1 & 0
\end{array}
\right] .
\end{equation}
Using (\ref{gm}), we match the value of the action for a fixed point
previously proposed by Keating \cite{keat1}, for the position representation 
\begin{equation}
S(q_{-}=q_f,q_{+}=q_f,{\bf m})=S(x_1,{\bf m}).
\end{equation}

In conclusion, the center and chord generating functions for
multidimensional cat maps are 
\begin{eqnarray}
S(x,{\bf m}) &=&xBx+x(B-{\frak J)}{\bf m}+\frac 14{\bf m}(B+\widetilde{%
{\frak J}}){\bf m}  \label{sx} \\
S(\xi ,{\bf m}) &=&\frac 14\xi \beta \xi +\frac 12\xi (\beta +{\frak J)}{\bf %
m}+\frac 14{\bf m}(\beta -\widetilde{{\frak J}}){\bf m.}  \label{scor}
\end{eqnarray}

The corresponding generating functions for the transformation $x_{+}={\cal M}x_{-}$
in the plane are just $S(x,0)$ and $S(\xi ,0)$. It is important to note that
all the reflections of the torus can be obtained with center points whose
coordinates are in $[0,\frac 12].$ It would thus be possible to define such
center points $x^{\prime }$ as 
\begin{equation}
x^{\prime }\equiv \frac{x_{+}+x_{-}}2\qquad \mbox{mod}(\frac 12).
\end{equation}
This choice would not lead to the explicit relation (\ref{xx-}) with the
winding number of the transformation, but instead 
\begin{equation}
x^{\prime }=\frac 12\left( {\cal M}+1\right) x_{-}-\frac{{\bf m}^{\prime }}2,
\label{xpm}
\end{equation}
where ${\bf m}^{\prime }$ has coordinates ${\bf m}_i^{\prime }=({\bf m}_i$
or ${\bf m}_i{\bf +1}),$ so that all the coordinates of $x^{\prime }$ would
be in $[0,\frac 12].$ Hence, we allow instead the center points $x,$ defined
as in (\ref{xdef}), to have coordinates in the full interval $[0,1],$
keeping the explicit relation with the winding number of the transformation.

As a consequence, the chords defined as in (\ref{cordef}),\ have coordinates
lying in the extended range $[-1,1].$ Thus, center points differing by
integer loops around the torus are equivalent and so are chords differing by
two integer loops.

\begin{eqnarray}
x &\equiv &x+{\bf k}  \label{xequi} \\
\xi &\equiv &\xi +2{\bf k.}  \label{corequi}
\end{eqnarray}
We find that (\ref{xequi}) and (\ref{corequi}) imply that, in (\ref{xcor})
and (\ref{corx}) respectively, the winding number ${\bf m}\,$ is equivalent
to: 
\begin{equation}
{\bf m}\equiv {\bf m}^{\prime }={\bf m}+({\cal M}-1){\bf k}  \label{mprim}
\end{equation}
in (\ref{xcor}) and 
\begin{equation}
{\bf m}\equiv {\bf m}^{\prime \prime }={\bf m}-({\cal M}+1){\bf k}.  \label{m2prim}
\end{equation}
in (\ref{corx}). This implies that replacing $\,$ ${\bf m}$ by ${\bf m}%
^{\prime }$ in the generating function $S(\xi ,{\bf m})$ will lead to
equivalent center points related by (\ref{xequi}). To obtain equivalent
chords related by (\ref{corequi}), it is necessary to replace ${\bf m}$ by $%
{\bf m}^{\prime \prime }\,$in the center generating function $S(x,{\bf m})$.
Performing the mentioned replacements we will obtain: 
\begin{eqnarray}
S(\xi ,{\bf m}^{\prime }) &=&S(\xi ,{\bf m})+\xi \wedge {\bf k-}\frac 12{\bf %
m}\Gamma _1{\bf k-}\frac 14{\bf k}\Delta _1{\bf k}  \label{scorper} \\
S(x,{\bf m}^{\prime \prime }) &=&S(x,{\bf m})-2x\wedge {\bf k-}\frac 12{\bf m%
}\Gamma _2{\bf k-}\frac 14{\bf k}\Delta _2{\bf k,}  \label{sxper}
\end{eqnarray}
where 
\begin{eqnarray}
\Gamma _1 &=&\left[ \left( {\frak J}+\widetilde{{\frak J}}\right) {\cal M}+\left( 
\widetilde{{\frak J}}-{\frak J}\right) \right]  \label{gama1} \\
\Delta _1 &=&\left[ \left( {\cal M}^t\widetilde{{\frak J}}{\cal M}+\widetilde{{\frak J}}%
\right) +{\cal M}^t\left( \widetilde{{\frak J}}-{\frak J}\right) +\left( {\frak J}+%
\widetilde{{\frak J}}\right) {\cal M}\right]  \label{delta1} \\
\Gamma _2 &=&\left[ \left( {\frak J-}\widetilde{{\frak J}}\right) {\cal M}+\left( 
{\frak J}+\widetilde{{\frak J}}\right) \right]  \label{gama2} \\
\Delta _2 &=&\left[ \left( {\cal M}^t\widetilde{{\frak J}}{\cal M}+\widetilde{{\frak J}}%
\right) -{\cal M}^t\left( {\frak J}+\widetilde{{\frak J}}\right) -\left( {\frak J}-%
\widetilde{{\frak J}}\right) {\cal M}\right] .  \label{delta2}
\end{eqnarray}

In this way we can restrict ${\bf m}$ to integer component vectors that lie
in one of the two fundamental parallelepipeds 
\begin{eqnarray}
&& 
\begin{array}{cc}
\Diamond _\xi =({\cal M}-1)\Box & \qquad \mbox{for }S(\xi ,{\bf m})
\end{array}
\label{para1} \\
&& 
\begin{array}{cc}
\Diamond _x=({\cal M}+1)\Box & \qquad \mbox{for }S(x,{\bf m})
\end{array}
\label{para2}
\end{eqnarray}
where $\Box $ is the unit hypercube that denotes the $2L$-torus. Hence, the
different orbits denoted by a given chord $\xi $ are given by all the
integer ${\bf m}$ lying in $\Diamond _\xi $. The number of such orbits is
independent of $\xi $, so taking $\xi =0$, we equate this to the number of
fixed points $\tau _\xi $, i.e. 
\begin{equation}
\tau ({\cal M})=|\det ({\cal M}-1)|=\frac{2^{2L}}{|\det ({\frak J}\beta -1)|}\equiv \tau
_\xi .  \label{d1}
\end{equation}
The different orbits that have the point $x$ as its center are denoted by
all the integers ${\bf m}$ lying now in $\Diamond _x.$ The number of these
orbits is given by the volume of $\Diamond _x$ which is 
\begin{equation}
\tau (-{\cal M})=\left| \det ({\cal M}+1)\right| =\frac{2^{2L}}{|\det ({\frak J}B+1)|}%
\equiv \tau _x.  \label{gama}
\end{equation}
Note that the number of periodic points of period two is $\tau _\xi \tau _x.$

For the matrix ${\cal M}$ to represent a $2L$-cat map it must be symplectic, so
that the map is area preserving, and ${\cal M}$ must have integer entries. Let us
now translate both the conditions for the $B$ and $\beta $ matrices. The
first condition implies that $B$ and $\beta $ are symmetric matrices, and
any symmetric matrix is associated through (\ref{mbbeta}) to a symplectic
matrix. The second condition restricts the $B$ and $\beta $ matrices to have
rational entries. Indeed, following (\ref{jbm}) and(\ref{jbetam}) we will
have 
\begin{eqnarray}
B &=&\frac{\overline{B}}{\det ({\cal M}+1)}\equiv \pm \frac{\overline{B}}{\tau _x}
\label{bbbar} \\
\beta &=&\frac{\overline{\beta }}{\det ({\cal M}-1)}=\pm \frac{\overline{\beta }}{%
\tau _\xi }  \label{betabar}
\end{eqnarray}
where $\overline{B}$ and $\overline{\beta }$ are symmetric matrices with
integer entries and the denominators are defined by (\ref{d1}) and (\ref
{gama}). It can happen that all the coefficients of the matrix $\overline{B}$
( or $\overline{\beta }$ ) have a common factor that is not coprime with $%
\tau _x$ (respectively $\tau _\xi $ ). Dividing by this common factor the
fraction in (\ref{bbbar}) ( or in (\ref{betabar}) ) is reduced to the form 
\begin{eqnarray}
B &=&\pm \frac{\overline{B}^{\prime }}{\tau _x^{\prime }}  \label{bbbar1} \\
\beta &=&\pm \frac{\overline{\beta }^{\prime }}{\tau _\xi ^{\prime }}.
\label{betabar1}
\end{eqnarray}
But not any symmetric $B$ or $\beta $ matrix with rational entries
guarantees that the associated symplectic matrix will have integer elements.
So we must find conditions on $B$ and $\beta $ for this to occur.

The characterization of the matrices $B$ or $\beta $ requires that the
corresponding transformation on the plane maps the points of an integer
lattice among themselves. We will examine the case $L=1$, as the extension
for many number of degrees of freedom follows easily. There are two
fundamental chords corresponding to fixed points on the torus: 
\begin{equation}
\xi _1=\left( 
\begin{array}{l}
1 \\ 
0
\end{array}
\right) \mbox{ \qquad and \qquad }\xi _2=\left( 
\begin{array}{l}
0 \\ 
1
\end{array}
\right) ,
\end{equation}
leading to the fixed points: 
\begin{equation}
x_j=\frac 12({\frak J}\beta + 1)\xi _{j\qquad ,}\mbox{ with }j=1,2.
\end{equation}
Of course, there is also $\xi _0=0$, but this ''plane fixed point'' makes no
restriction on the torus map. For the transformation to be a cat map, all
the corners of the fundamental parallelogram $\Box $ must be fixed points,
so, for any integers $r$ and $s$, there are integers $m_1$ and $m_2$ such
that: 
\begin{equation}
m_1x_1+m_2x_2=\left( 
\begin{array}{l}
r \\ 
s
\end{array}
\right) =\frac 12({\frak J}\beta + 1)\left( 
\begin{array}{l}
{m_1} \\ 
{m_2}
\end{array}
\right) .
\end{equation}
This is true if only if $2({\frak J}\beta + 1)^{-1}$ has integer entries. This condition is general for any degrees of freedom. Similarly, we find that $2({\frak J}\beta - 1)^{-1}$ having integer entries is also a necessary and sufficient condition for the corresponding ${\cal M}^{-1} $ to define a cat map. 
But , if ${\cal M}$ defines a cat map, so does $-{\cal M}$ with the associated chord
matrix $-B.$ Therefore, it is also a necessary and sufficient condition, for
a center generating function to determine a cat map, that the associated
center matrix $B$ have the property that $2({\frak J}B\pm 1)^{-1}$ be an
integer matrix. Evidently we easily find a subclass of cat maps by restricting their Cayley parametrization to $B$ (or $ \beta$ ) matrices of the form (\ref{bbbar}) such that $
det(1\pm{\frak J}B)=\pm 1, \pm 2
$, then $2(1+{\frak J}B)^{-1}$ is an integer matrix.

Although the conditions on the symmetric matrices $B$ or $\beta \,\,$to
denote a cat map are not as trivial as the ones on the symplectic matrix ${\cal M},$
it is simpler to find rational symmetric matrices that fulfill the condition
on $\beta $ and $B$ than to find integer symplectic matrices. The fact that $%
B$ or $\beta $ are symmetric and of the form (\ref{bbbar}) allows us to find cat maps by sampling  $\left[ (L)\times (2L+1)+1\right] $ integer numbers. Otherwise,
to fulfill the condition (\ref{msimple}), needs a loop over $(2L)^2$ integer
coefficients.

To conclude this section, we verify the property of feline invariance for
the chord and center generating functions. First, we note that, symplectic
invariance in the plane \cite{ozrep} implies that under a symplectic
coordinate transformation $x\rightarrow x^{\prime }={\cal N}x,$ $S(x,0)=S(x^{\prime
},0)$ and $S(\xi ,0)=S(\xi ^{\prime },0),$ with $\xi ^{\prime }={\cal N}\xi .$ But
it is also evident that the winding number ${\bf m}$ transforms in the same
manner: ${\bf m}^{\prime }={\cal N}{\bf m.}$ As far as the $x$ dependent term in $%
S(x,{\bf m}),$ we thus find that the effect of the feline transformation is
merely that of substituting $B\rightarrow {\cal N}^tB{\cal N},$ and similarly the change
in $S(\xi ,{\bf m})$ is obtained from $\beta \rightarrow {\cal N}^t\beta {\cal N}.$ The
constant terms $f({\bf m})$ and $g({\bf m})$ in (\ref{sxf}) and (\ref{scorg}%
) are not invariant under a feline transformation in the form (\ref{gm})
that we have chosen to match reference \cite{keat1}, so that it is
preferable to use (\ref{3.30}) and (\ref{3.30g}) when dealing with
equivalence classes of cat maps.

It is important to note that, unlike the symmetric matrices $B$ and $\beta $%
, ${\frak J}B\rightarrow {\cal N}^{-1}{\frak J}B{\cal N}$ and ${\frak J}\beta \rightarrow
{\cal N}^{-1}{\frak J}\beta {\cal N}$, under a similarity transformation ${\cal M}\rightarrow
{\cal N}^{-1}{\cal M}{\cal N}.$ Therefor, the eigenvalues of ${\frak J}B$ and ${\frak J}\beta $
are feline invariant, just as those of ${\cal M}$, and can thus be used to classify
cat maps.

\section{Classification of classical cat maps}

The periodic orbits of cat maps have been studied in great details by
Percival and Vivaldi \cite{parcivivaldi} and also by Keating in \cite{keat1}
for one degree of freedom and their results were recently extended to an
arbitrary number of degrees of freedom \cite{greenman}. It is shown that a
point on the unit $2L$-torus is periodic if and only if all its coordinates
are rational and any grid of points with rational coordinates is invariant
under the action of the map. From (\ref{mapa}) we can see that the periodic
points $x_l$ of integer period $l$ are labeled by the winding numbers ${\bf %
m,}$ so that 
\begin{equation}
x_l=\left( 
\begin{array}{l}
{p_l} \\ 
{q_l}
\end{array}
\right) =({\cal M}^l-1)^{-1}{\bf m=}({\frak J}\beta ^{\left( l\right) }-1)\frac{%
{\bf m}}2.  \label{xfix}
\end{equation}
Here $\beta ^{\left( l\right) }$ denotes the symmetric matrix associated to $%
{\cal M}^l$ through (\ref{jbetam}). To have $x_l$ on the unit $2L$-hypercube $\Box $%
, ${\bf m}$ must lie within the parallelepiped formed by the action of the
matrix $({\cal M}^l-1)$ on $\Box .$ Hence, the number of integer points ${\bf m}$
is given by its hypervolume, so that the number of periodic points with
period $l$ is 
\begin{equation}
\tau ({\cal M}^l)=|\det ({\cal M}^l-1)|=\frac{2^{2L}}{|\det ({\frak J}\beta ^{\left(
l\right) }-1)|}.  \label{npfix}
\end{equation}
According to (\ref{xfix}) the periodic points of period $l$ form a lattice
in phase space with rational coordinates.

The motion of any point $x_{-}=x_1+\delta _{-}$ near a fixed point $x_1$
will be 
\begin{equation}
{\cal M}x_{-}={\cal M}(x_1+\delta _{-})=x_1+{\cal M}\delta _{-}=x_1+\delta _{+}=x_{+}.
\end{equation}
To determine the character of such a motion we have to study the eigenvalues
and eigenvectors of the matrix ${\cal M}$: $\,$ 
\begin{equation}
\lambda _{\cal M}^k=\left| \lambda _{\cal M}^k\right| e^{i\theta _k}
\end{equation}
The modulus $\left| \lambda _{\cal M}^k\right| $ indicates that the motion is
stretching while the argument $e^{i\theta _k}$ indicates rotation around the
fixed point $x_1.$ For a symplectic matrix ${\cal M}$, if $\lambda _{\cal M}$ is an
eigenvalue of ${\cal M}$ then $\lambda _{\cal M}^{*}$, $\frac 1{\lambda _{\cal M}}$ and $\frac
1{\lambda _{\cal M}^{*}}$ will also be eigenvalues of ${\cal M}.$

The classification of the eigenvalues is possible in either the ${\cal M},B,$ or $%
\beta $ descriptions. Using (\ref{mbbeta}) we obtain the relation with the
eigenvalues of ${\frak J}B$ and ${\frak J}\beta $ denoted respectively as $%
\lambda _{{\frak J}B}$ and $\lambda _{{\frak J}\beta },$%
\begin{equation}
\lambda _{\cal M}=\frac{\left( 1-\lambda _{{\frak J}B}\right) }{\left( 1+\lambda _{%
{\frak J}B}\right) }=\frac{\left( \lambda _{{\frak J}\beta }+1\right) }{%
\left( \lambda _{{\frak J}\beta }-1\right) }  \label{lmbbeta}
\end{equation}
and inversely: 
\begin{eqnarray}
\lambda _{{\frak J}B} &=&\frac{\left( 1-\lambda _{\cal M}\right) }{\left( 1+\lambda
_{\cal M}\right) }  \label{ljbm} \\
\lambda _{{\frak J}\beta } &=&\frac{\left( \lambda _{\cal M}+1\right) }{\left(
\lambda _{\cal M}-1\right) }=-\frac 1{\lambda _{{\frak J}B}}.  \label{ljbetam}
\end{eqnarray}
In this way, if $\lambda _{{\frak J}B}$ is an eigenvalue of ${\frak J}B,$
then $\lambda _{{\frak J}B}^{*}$, $-\lambda _{{\frak J}B}$ and $-\lambda _{%
{\frak J}B}^{*}$ will also be. In the same way, if $\lambda _{{\frak J}\beta
}$ is an eigenvalue of ${\frak J}\beta ,$ then $\lambda _{{\frak J}\beta
}^{*}$, $-\lambda _{{\frak J}\beta }$ and $-\lambda _{{\frak J}\beta }^{*}$
also are.

For cat maps with $L=2$ the matrices ${\cal M},B,$ and $\beta $ will be $4\times 4$
and we then have the following generic cases for the eigenvalues

\begin{enumerate}
\item  Elliptic: there is a conjugate pair of $\lambda _{\cal M}$ both on the unit
circle and conjugate pairs of purely imaginary $\lambda _{{\frak J}B}$ and $%
\lambda _{{\frak J}\beta }.$

\item  Hyperbolic: there is a pair $(\lambda _{\cal M},\frac 1{\lambda _{\cal M}})$ on the
real axis and pairs $(\lambda _{{\frak J}B},-\lambda _{{\frak J}B})$ and $%
(\lambda _{{\frak J}\beta },-\lambda _{{\frak J}\beta })$ on the real axis.

\item  Parabolic: there are degenerate eigenvalues $\lambda _{\cal M}=\pm 1$. Then
for $\lambda _{\cal M}=1,$ $\beta $ is singular and $\lambda _{{\frak J}B}=0$; for $%
\lambda _{\cal M}=-1,$ $B$ is singular and $\lambda _{{\frak J}\beta }=0$.

\item  Mixed: each pair belongs to a different one of the above categories.

\item  Loxodromic: the eigenvalues of ${\cal M},B,$ or $\beta $ matrices form
quartets of complex eigenvectors of the form $(\lambda _{\cal M},\frac 1{\lambda
_{\cal M}},\lambda _{\cal M}^{*},\frac 1{\lambda _{\cal M}^{*}})\,$ for the ${\cal M}$ matrix and $%
(\lambda _{{\frak J}S},-\lambda _{{\frak J}S},\lambda _{{\frak J}%
S}^{*},-\lambda _{{\frak J}S}^{*})\,$ for $S$ being one of the symmetric
matrices $B$ or $\beta .$
\end{enumerate}

The first three cases arise also for one degree of freedom cat maps. The
case $L=2$ is the lowest number of degrees of freedom where not only
elliptic, parabolic and hyperbolic fixed points will appear, but also
loxodromic ones. For more degrees of freedom no new cases will occur; there
is only a greater variety of mixed cases. Nongeneric possibilities arise for
any dimension for continuous families of systems \cite{quinze}, but we do
not know if there exists any corresponding cat map.

In the general case, the motion is the composition of stretching and
rotation, but if the angles of the rotation are of the form $\theta _k=\frac
ij2\pi ,$ after $j\,$ applications of the map, the matrix ${\cal M}^j$ has only
real eigenvalues that come in pairs $\left| \lambda _{\cal M}^k\right| ^j$ and $%
\left| \lambda _{\cal M}^k\right| ^{-j}$. The dynamics has then an ignorable
coordinate; the angles are constants of the motion for ${\cal M}^j.$ For one degree
of freedom systems, where there are only two eigenvalues, $\lambda _{\cal M}$ and $%
\lambda _{\cal M}^{*},$\ the condition 
\begin{equation}
Tr\left( {\cal M}\right) =\lambda _{\cal M}+\lambda _{\cal M}^{*}=\mbox{integer}
\end{equation}
implies that in the elliptic case only angles of the form $\theta _k=\frac
ij\pi $ with $j=2$ or $3$ are allowed. This is an example of a rational
restriction on $\theta _k$, that is, ${\cal M}^j$ will be the identity map. This
result is in accordance with Ma\~{n}e's theorem \cite{mane} that
two-dimensional symplectomorphisms are either Anosov or they have zero
entropy. Nonetheless, irrational rotation angles do exist in the loxodromic
or mixed cases for $L>1$.

We now turn our attention to the classification of four-dimensional cat
maps. Recalling that the characteristic polynomial for any $k\times k$
matrix $A$ is

\begin{equation}
P_A(\lambda )=\det (A-\lambda )=\sum_{n=1}^k\alpha _n\lambda ^{k-n},
\end{equation}
where the $\alpha _n$ coefficients are given by the recurrence relation, 
\begin{eqnarray}
\alpha _0 &=&1, \\
\alpha _n &=&-\frac 1n\sum_{i=1}^n\alpha _{n-i}a_i\qquad \mbox{ with }%
a_i=Tr(A^i),
\end{eqnarray}
we obtain that for any symmetric matrix $B$ (i.e. $B$ and $\beta $ ), 
\begin{equation}
P_{{\frak J}B}(\lambda )=P_{{\frak J}B}(-\lambda )  \label{psym}
\end{equation}
so that 
\begin{equation}
P_{{\frak J}B}(\lambda )=\lambda ^4-\frac 12b_2\lambda ^2+\det B  \label{pbl}
\end{equation}
with $b_2=Tr\left[ \left( {\frak J}B\right) ^2\right] ,($ note that $\det 
{\frak J}B=\det B)$. There is a similar expression for $P_{{\frak J}\beta
}(\lambda )$, required when the center representation is singular, 
\begin{equation}
P_{{\frak J}\beta }(\lambda )=\lambda ^4-\frac 12\beta _2\lambda ^2+\det
\beta .  \label{pbetal}
\end{equation}
where $\beta _2=Tr\left[ \left( {\frak J}\beta \right) ^2\right] .$ For the
symplectic case, we obtain 
\begin{equation}
P_{\cal M}(\lambda )=\lambda ^4-Tr({\cal M})\lambda +\frac 12\left[ Tr({\cal M}^2)-Tr({\cal M})^2\right]
\lambda ^2-Tr({\cal M})\lambda ^3+1  \label{pml}
\end{equation}
which is harder to analyze, so we will perform the classification of the
different behaviors using ${\frak J}B$ or ${\frak J}\beta $. Solving $P_{%
{\frak J}B}(\lambda )=0$ using (\ref{pbl}), leads to 
\begin{equation}
\lambda _{{\frak J}B}=\pm \sqrt{\frac{b_2}4\pm \sqrt{\left( \frac{b_2}%
4\right) ^2-\det B}},
\end{equation}
whereas 
\begin{equation}
\lambda _{{\frak J}\beta }=\pm \sqrt{\frac{\beta _2}4\pm \sqrt{\left( \frac{%
\beta _2}4\right) ^2-\det \beta }}.
\end{equation}
We can now classify the different behaviors according to the feline
invariants of the matrix ${\frak J}B\,$ or ${\frak J}\beta $, the
eigenvalues of the symplectic matrix ${\cal M}$ being obtained with the help of (%
\ref{lmbbeta}). The loxodromic behavior corresponding to four complex
eigenvalues will appear if the square root term $\left[ \left( \frac{b_2}%
4\right) ^2-\det B\right] $ inside the square root is negative. That is, the
loxodromic behavior will appear only if 
\begin{equation}
\det B>\left( \frac{b_2}4\right) ^2\qquad \mbox{ or equivalently }\qquad
\det \beta >\left( \frac{\beta _2}4\right) ^2.  \label{loxcond}
\end{equation}

\begin{figure}[tbp]
\centerline {\epsfxsize=5in \epsffile{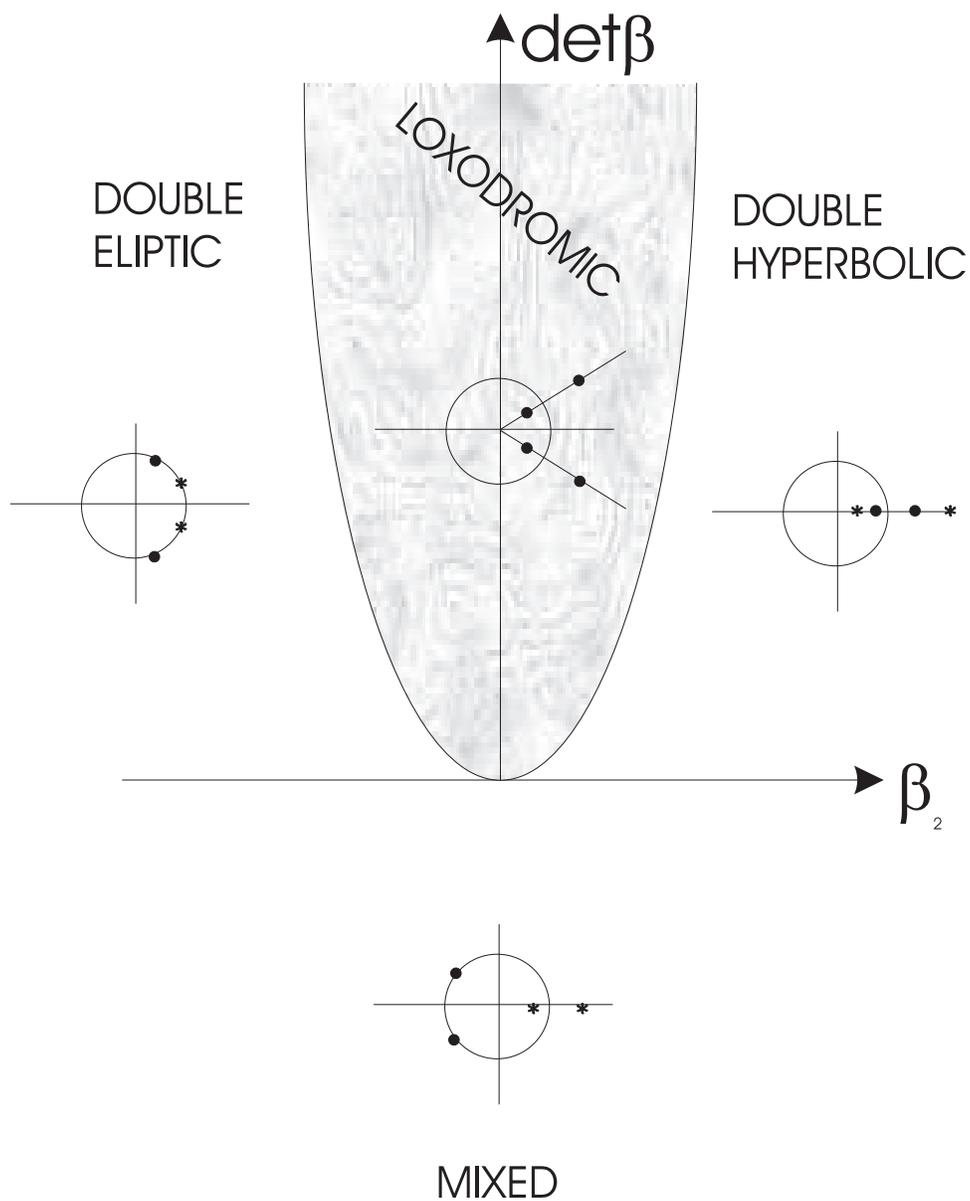} }
\caption{ A classification of the different cat maps for two degrees of
freedom according to the feline invariants of the ${\frak J}\beta $ matrix.
The parabolic boundary is $\det \beta =\left( \frac{\beta _2}4\right) ^2.$ A
similar picture exist for the ${\frak J}B$ matrix. }
\label{fig.1}
\end{figure}

In fig~\ref{fig.1} we find a complete classification of the different
types of behavior according to the feline invariants of ${\frak J}\beta $
and the same arises for the invariants of ${\frak J}B.$ These invariants
also allow us to obtain $\tau _\xi ,\,$the number of orbits for any chord $%
\xi ,$%
\begin{equation}
\tau _\xi =\left| \frac{2^4}{P_{{\frak J}\beta }(1)}\right| =\left| \frac{%
2^4\det B}{P_{{\frak J}B}(1)}\right| ,
\end{equation}
or $\tau _x$ the number of orbits centered on $x$ as 
\begin{equation}
\tau _x=\left| \frac{2^4}{P_{{\frak J}B}(1)}\right| =\left| \frac{2^4\det
\beta }{P_{{\frak J}\beta }(1)}\right| .
\end{equation}

\subsection*{Examples}

We here show some examples of different types of cat maps. The examples are
such that the $\beta $ matrix, that denotes the chord representation, has
integer elements.

\subsubsection*{ The Hannay and Berry Cat}

It is important to see that the first cat map to be quantized ( the Hannay
and Berry cat map) \cite{hanay} can also be treated with the formalism
described here . For a $2\times 2$ matrix, the characteristic polynomial
reads, 
\begin{equation}
P_{{\frak J}\beta }(\lambda )=\lambda ^2-\frac 12\beta _2.
\end{equation}
For the symplectic matrix 
\begin{equation}
{\cal M}_{hb}=\left[ 
\begin{array}{cc}
2 & 1 \\ 
3 & 2
\end{array}
\right] \mbox{, with  the associated symmetric matrix }\beta _{hb}=\left[ 
\begin{array}{cc}
3 & 0 \\ 
0 & -1
\end{array}
\right] ,  \label{mhb}
\end{equation}
so that 
\[
\beta _2=Tr\left[ \left( {\frak J}\beta _{hb}\right) ^2\right] =6,
\]
the number of fixed points is 
\[
\tau _\xi ^{hb}=\left| \frac{2^2}{P_{{\frak J}\beta }(\lambda )}\right| =2
\]
and the eigenvalues of the matrix ${\cal M}_{hb}$ are 
\begin{eqnarray*}
\lambda _1^{hb} &=&2+\sqrt{3} \\
\lambda _2^{hb} &=&2-\sqrt{3}.
\end{eqnarray*}
The map is then hyperbolic.

\subsubsection*{ The double hyperbolic case}

Let us now study cat maps that have two pairs of eigenvalues both in the
real axis, that is, maps belonging to the first quadrant in figure \ref{fig.1}%
. The $\beta _{hh}\,$ matrix below describes this case, the associated
symplectic matrix being ${\cal M}_{hh}$ : 
\begin{equation}
\beta _{hh}=\left[ 
\begin{array}{cccc}
0 & 2 & 1 & 2 \\ 
2 & 0 & 2 & 1 \\ 
1 & 2 & 0 & 0 \\ 
2 & 1 & 0 & 1
\end{array}
\right] \quad \mbox{and }\quad {\cal M}_{hh}=\left[ 
\begin{array}{cccc}
2 & -2 & -1 & 0 \\ 
-2 & 3 & 1 & 0 \\ 
-1 & 2 & 2 & 1 \\ 
2 & -2 & 0 & 1
\end{array}
\right] .  \label{mhh}
\end{equation}
The invariants are 
\[
\det \beta _{hh}=17\quad \mbox{and }\quad \beta _2=Tr\left[ \left( {\frak J}%
\beta _{hh}\right) ^2\right] =20. 
\]
Thus, the number of fixed points $\tau _\xi ^{hh}=2$ and the eigenvalues of
the symplectic matrix ${\cal M}$ are 
\begin{eqnarray*}
\lambda _1^{hh} &=&2.112388 \\
\lambda _2^{hh} &=&0.4739 \\
\lambda _3^{hh} &=&5.22 \\
\lambda _4^{hh} &=&0.1914.
\end{eqnarray*}

\subsubsection*{ The mixed case}

We consider here the case where the eigenvalues of the map are of mixed
nature, i.e. there is a couple of real eigenvalues and a complex conjugated
pair on the unit circle. This corresponds to $\beta $ matrices that belong
to the third and fourth quadrant in figure ~\ref{fig.1}.

One matrix of this kind is $\beta _{eh1}$, the associated symplectic matrix
being ${\cal M}_{eh1}:$ 
\begin{equation}
\beta _{eh1}=\left[ 
\begin{array}{cccc}
0 & 2 & 1 & 2 \\ 
2 & 0 & 2 & 1 \\ 
1 & 2 & 0 & 2 \\ 
2 & 1 & 2 & 0
\end{array}
\right] \quad \mbox{and }\quad {\cal M}_{eh1}=\left[ 
\begin{array}{cccc}
0 & 0 & -1 & 0 \\ 
0 & 0 & 0 & -1 \\ 
1 & 0 & 2 & 1 \\ 
0 & 1 & 1 & 2
\end{array}
\right] .  \label{meh1}
\end{equation}
The invariants are 
\[
\det \beta _{eh1}=-15\quad \mbox{and }\quad \beta _2=Tr\left[ \left( {\frak J%
}\beta _{eh1}\right) ^2\right] =4. 
\]
Thus, the number of fixed points $\tau _\xi ^{eh1}=1$ and the eigenvalues of
the symplectic matrix ${\cal M}$ are 
\begin{eqnarray*}
\lambda _1^{eh1} &=&2.6180 \\
\lambda _2^{eh1} &=&0.381966 \\
\lambda _3^{eh1} &=&\exp \left( i\frac{2\pi }6\right) \\
\lambda _4^{eh1} &=&\exp \left( -i\frac{2\pi }6\right) .
\end{eqnarray*}
This example shows rotation angles that are fractional multiples of $\pi $,
for which the dynamics are equivalent to that of hyperbolic systems with one
degree of freedom . Another example of a mixed system is given by the matrix 
$\beta _{eh2}$, the associated symplectic matrix being ${\cal M}_{eh2}:$ 
\begin{equation}
\beta _{eh2}=\left[ 
\begin{array}{cccc}
0 & 2 & 1 & 2 \\ 
2 & 0 & 2 & 1 \\ 
1 & 2 & 0 & 2 \\ 
2 & 1 & 2 & 1
\end{array}
\right] \quad \mbox{and }\quad {\cal M}_{eh2}=\left[ 
\begin{array}{cccc}
0 & 0 & -1 & 0 \\ 
0 & -1 & -1 & -2 \\ 
1 & 0 & 2 & 1 \\ 
0 & 2 & 2 & 3
\end{array}
\right] .  \label{meh2}
\end{equation}
Now we have 
\begin{equation}
\det \beta _{eh2}=-7\quad \mbox{and }\quad \beta _2=Tr\left[ \left( {\frak J}%
\beta _{eh2}\right) ^2\right] =4.
\end{equation}
Thus, the number of fixed points $\tau _\xi ^{eh2}=2$ and the eigenvalues of
the symplectic matrix ${\cal M}$ are 
\begin{eqnarray*}
\lambda _1^{eh2} &=&3.0906578 \\
\lambda _2^{eh2} &=&0.32355571 \\
\lambda _3^{eh2} &=&\exp \left( i1.27354496\right) \\
\lambda _4^{eh2} &=&\exp \left( -i1.27354496\right) .
\end{eqnarray*}
In this example there are no rotations with angles that are a fraction of $%
\pi ,$ the dynamics will then be ergodic and mixing in the whole phase space.

\subsubsection*{ The loxodromic case}

We choose now a $\beta $ matrix that belongs to the loxodromic region in Fig
~\ref{fig.1}, for example $\beta _{lox}$, whose associated symplectic
matrix being ${\cal M}_{eh1}:$ 
\begin{equation}
\beta _{lox}=\left[ 
\begin{array}{cccc}
0 & 2 & 1 & 0 \\ 
2 & 0 & 2 & 1 \\ 
1 & 2 & 1 & 1 \\ 
0 & 1 & 1 & 1
\end{array}
\right] \quad \mbox{and }\quad {\cal M}_{lox}=\left[ 
\begin{array}{cccc}
0 & 1 & 0 & 0 \\ 
0 & 1 & 1 & 0 \\ 
1 & -1 & 1 & 1 \\ 
-1 & -1 & -2 & 0
\end{array}
\right]  \label{mlox}
\end{equation}
whose invariants are

\begin{equation}
\det \beta _{lox}=5\quad \mbox{and }\quad \beta _2=Tr\left[ \left( {\frak J}%
\beta _{lox}\right) ^2\right] =-4.  \nonumber
\end{equation}
Thus, the number of fixed points $\tau _\xi ^{lox}=2$ and the eigenvalues of
the symplectic matrix ${\cal M}$ are 
\begin{eqnarray*}
\lambda _1^{lox} &=&1.7000157\exp \left( i1.1185178\right) \\
\lambda _2^{lox} &=&1.7000157\exp \left( -i1.1185178\right) \\
\lambda _3^{lox} &=&0.5882298\exp \left( i1.1185178\right) \\
\lambda _4^{lox} &=&0.5882298\exp \left( -i1.1185178\right) .
\end{eqnarray*}
There are no rational rotations, so the dynamics is ergodic and mixing in
the whole phase space.

\subsubsection*{The double elliptic case}

In this case, the eigenvalues are characterized by two rotation angles. An
interesting feature is that the condition that the ${\cal M}$ matrix have integer
elements, prevents either angle from being an irrational multiple of $\pi .$
This restricts the dynamics to the trivial integrability observed for one
degree of freedom systems.

In conclusion, any symplectomorphism on the $4$-torus belongs to one of the
following cases:

\begin{enumerate}
\item  Ergodic on the full phase space (a manifold of 2 degrees of freedom).
For the double hyperbolic, mixed or loxodromic cases with irrational
rotation angles.

\item  Ergodic in a $2$-dimensional manifold . We have a constant of the
motion, then the ergodicity is restricted to a low dimensional manifold (a
manifold with $L=1).$ For the mixed or loxodromic cases with rational
rotation angles.

\item  A root of unity: that is, we have two constants of the motion. All
orbits are periodic and have the same period for this double elliptic cases.
\end{enumerate}

\section{Simple Quantum Cat maps}

\setcounter{equation}{0}

Quantum dynamics is characterized by a unitary evolution operator, or
propagator. It is possible to obtain the center and chord representation of
an operator on the torus from its counterpart on the plane. This will be the
way that we quantize cat maps, leaving the general construction for the next
section. In some cases the quantum propagator denoting cat maps in the
center or chord representation acquires simple expressions; in this section
we will treat these special cases. More details about torus Hilbert space
and its center and chord representations based on reference \cite{opetor}
are available in Appendix A.

It is well known that torus quantization implies that the Hilbert space $%
\left[ {\cal H}_N^\chi \right] ^L$ associated to the $2L$-torus has finite
dimension $N^L$ and is characterized by a vector Floquet parameter $\chi
=(\chi _p,\chi _q)$ whose components are real numbers belonging to $\left[
0,1\right] .$ The fact that $\left[ {\cal H}_N^\chi \right] ^L$ has finite
dimension, implies that position and momentum eigenstates can only take on a
set of discrete values that form a discrete lattice called {\it quantum
phase space} (QPS). Any point $x$ in this QPS has coordinates, 
\begin{equation}
x=\left( 
\begin{array}{c}
p_m \\ 
q_n
\end{array}
\right) =\frac 1N\left( 
\begin{array}{c}
m+\chi _p \\ 
n+\chi _q
\end{array}
\right) .
\end{equation}

The center and chord representations are based on translation and reflection
operators on this QPS as we show in \cite{opetor}. Chords are of the form 
\begin{equation}
\xi _{r,s}=\frac 1N\left( 
\begin{array}{c}
r \\ 
s
\end{array}
\right) =\frac 1N\bar{\xi},
\end{equation}
with $r$ and $s$ integer numbers and $\bar{\xi}=\left( 
\begin{array}{c}
r \\ 
s
\end{array}
\right) $, while the center points $x_{a,b}$ are labeled by half-integer
numbers $a$ and $b,$ 
\begin{equation}
x_{a,b}=\frac 1N\left( 
\begin{array}{c}
a+\chi _p \\ 
b+\chi _q
\end{array}
\right) .
\end{equation}
At a first stage we restrict our attention to those maps that can be
quantized on the Floquet parameters $\chi =(0,0).$ As we will see in section
5 this implies that the matrix ${\cal M}$ must satisfy 
\begin{equation}
\sum_{j=1}^L{\cal M}_{i,j}{\cal M}_{i,j+L}=\mbox{even integer for all }i.  \label{Mquanti1}
\end{equation}

As we will show in section 5, if $2\tau _\xi ^{\prime }$, defined in (\ref
{betabar1}), and $N$ are coprime numbers, a complete representation of the
propagator is obtained having the symbol on a lattice of chords $\Xi $ such
that 
\begin{equation}
\Xi =\xi +{\bf n=}\frac{2\tau _\xi ^{\prime }}N\overline{\Xi },
\label{corN1}
\end{equation}
where the components of $\overline{\Xi }$ are integer numbers up to $N.$ So
for any chord $\xi $ there is an equivalent chord $\Xi .$ For the allowed
values of $N$, the propagator for cat maps in the chord representation takes
the form 
\begin{equation}
{\bf U}_{\cal M}(\Xi )=2^L\left( \tau _\xi \right) ^{-\frac 32}e^{-i2\pi N\left[
\frac 14\Xi \beta \Xi \right] }\sum_{{\bf m\in \Diamond }_\xi }e^{-i2\pi
N\frac 14{\bf m}(\beta -\widetilde{{\frak J}}){\bf m}},  \label{ugcorimp}
\end{equation}
where $\tau _\xi $ is the number of fixed points of the classical map. For $%
\beta $ matrices that fulfill the feline conditions, the symbol ${\bf U}%
_{\cal M}^0(\Xi )$ must represent an unitary operator. In that case (\ref{unitcor})
shows that it must has the form 
\begin{equation}
{\bf U}_{\cal M}(\Xi )=\frac{e^{i\varphi _N({\cal M})}}{\sqrt{N}^L}e^{-i2\pi N\left[ \frac
14\Xi \beta \Xi \right] },
\end{equation}
which restricts 
\begin{equation}
\frac{e^{i\varphi _N({\cal M})}}{\sqrt{N}^L}=2^L\left( \tau _\xi \right) ^{-\frac
32}\sum_{{\bf m\in \Diamond }_\xi }e^{-i2\pi N\frac 14{\bf m}(\beta -%
\widetilde{{\frak J}}){\bf m}}.
\end{equation}
At first sight, the phase $\varphi _N({\cal M})$ is only an unimportant global
phase factor, but the interference of the different $\varphi _N({\cal M}^l)$ for
the different powers $l$ of the map will have a crucial importance for the
density of states.

As $\xi $ and $\Xi $ are equivalent chords, the symbol ${\bf U}_{\cal M}(\xi )\,$
and ${\bf U}_{\cal M}(\Xi )$ are related by symmetry relations (\ref{eq:Acorsim})
so that 
\begin{equation}
{\bf U}_{\cal M}(\xi )=\frac{e^{i\varphi _N({\cal M})}}{\sqrt{N}^L}e^{-i2\pi N\left[ S(\xi
,{\bf n})\right] },  \label{ucors}
\end{equation}
where $S(\xi ,{\bf n})$ is the action of the classical orbit whose chord is $%
\xi $ and that performs ${\bf n}$ loops around the torus as defined in (\ref
{scor}).

If we chose the symplectically invariant form (\ref{3.30g}) for the $\xi $%
-independent part of the chord generating function, instead of (\ref{gm}),
we would have in (\ref{ucors}) a supplementary phase factor $e^{i2\pi N\frac
14{\bf n}\widetilde{{\frak J}}{\bf n}}=e^{i\gamma _{{\bf n}}}$ with $\gamma
_{{\bf n}}$ a "Maslov index" for the orbit. This observation is true for
all the following quantum theory.

In the case of the center representation, for  $2\tau _x^{\prime }\,,$
defined in (\ref{bbbar1}), and $N$ coprime numbers, a complete representation of the
propagator is obtained by performing a transformation to center points $X$
that are integer multiples of $\frac{\tau _x^{\prime }}N$, 
\begin{equation}
X=x+\frac 12{\bf j=}\frac{\tau _x^{\prime }}N\overline{X},  \label{xN1}
\end{equation}
where the components of $\overline{X}$ are integer numbers up to $N$ \cite
{opetor}$.$ On these points the center representation of the propagator
takes the form 
\begin{eqnarray}
{\bf U}_{\cal M}(X) &=&2^L\tau _x^{-\frac 32}e^{i2\pi N\left[ XBX\right] }\sum_{%
{\bf m\in \Diamond }_x}e^{i2\pi N\frac 14{\bf m}(B+\widetilde{{\frak J}})%
{\bf m}}  \label{ugxd} \\
&=&e^{i\varphi _N^{\prime }({\cal M})}e^{i2\pi N\left[ XBX\right] },
\end{eqnarray}
where the last equality is obtained by imposing the unitarity of ${\bf \hat{U%
}}_{\cal M}$ and using (\ref{unitx}). Hence, we define the angle $\varphi
_N^{\prime }({\cal M})$ so 
\begin{equation}
e^{i\varphi _N^{\prime }({\cal M})}=2^L\tau _x^{-\frac 32}\sum_{{\bf m\in \Diamond }%
_x}e^{i2\pi N\frac 14{\bf m}(B+\widetilde{{\frak J}}){\bf m}}.
\end{equation}
From the symmetry relations (\ref{eq:Acensim}), we find that the symbols on
the original points $x$ are 
\begin{equation}
{\bf U}_{\cal M}(x)=e^{i\varphi _N^{\prime }({\cal M})}e^{i2\pi N\left[ S(x,{\bf j}%
)\right] },
\end{equation}
where here $S(x,{\bf j})\,$ is the center generating function, defined in (%
\ref{sx}), on a center point $x$ for an orbit performing ${\bf j}$ loops.
The cases above are then special cases where the propagator on the torus has
the same form as its equivalent on the plane; thus, they are ideally suited
for the comparison of classical and quantum motion.

To obtain the more familiar position representation of the propagator from
its chord representation, we use (\ref{eq:AQcor}), 
\begin{equation}
{\bf U}_{\cal M}{\bf (q}_m,{\bf q}_n)=\frac{e^{i\varphi _N({\cal M})}}{(N)^{\frac{3L}2}}%
\sum_{\xi _p=0}^{N-1}\exp \left\{ -i2\pi N\left[ S(\xi _{p,m-n},{\bf n)+}%
\frac{q_m+q_n}2\xi _p\right] \right\} ,
\end{equation}
While (\ref{eq:AQx}) allows us to obtain the position representation from
the center one: 
\begin{equation}
{\bf U}_{\cal M}{\bf (q}_m,{\bf q}_n)=\frac{e^{i\varphi _N^{\prime }({\cal M})}}{N^L}%
\sum_{x_p=0}^{\frac{N-1}2}\exp \left\{ i2\pi N\left[ S(x_{p,\frac{m+n}2},%
{\bf j)+}\left( q_m-q_n\right) x_p\right] \right\} .  \label{eq:UQx}
\end{equation}
For the case of one degree of freedom, this leads to the Hannay and Berry
propagator.

As we will see in section 5, condition (\ref{Mquanti1})\ is preserved for
the different powers of the map, hence, if a map is quantizable on the
Floquet parameters $\chi =(0,0),$ all its powers also are. Furthermore, the
propagator for ${\cal M}^l$ has the form 
\begin{equation}
{\bf U}_{\cal M}(\Xi )=\frac{e^{i\varphi _N({\cal M})}}{\sqrt{N}^L}e^{-i2\pi N\left[ \frac
14\Xi \beta ^{(l)}\Xi \right] },  \label{ugcorl}
\end{equation}
in the chord representation, where $\beta ^{(l)}$ denotes the symmetric
matrix associated to ${\cal M}^l$ through (\ref{jbetam}) and $\Xi $ are points on a
lattice of the form 
\begin{equation}
\Xi {\bf =}\frac{\tau ({\cal M}^l)}N\overline{\Xi }.
\end{equation}

As we discussed in section 2, lattices of rational points are invariant
under classical cat maps. The denominator $g$ of these rational points can
then be used to label the lattice, i.e. there exists a minimal period $l_g$
under which all the points on the lattice are fixed for ${\cal M}^{l_g}.\,$In other
words, ${\cal M}^{l_g},$ restricted to the mentioned lattice, is the identity. We
will call $l_g$ the classical periodicity function of the lattice. As we can
see in the appendix, torus quantization is performed on a lattice of
rational points whose denominator is $N$. Under these considerations we find
that the quantum propagator is also periodic.

It is now possible to define conditions on the matrix ${\cal M}$ and the dimension
of the Hilbert space $N\,$ under which $\widehat{{\bf U}}_{\cal M}\,$ reduces to
the identity operator, i.e. 
\begin{equation}
\widehat{{\bf U}}_{\cal M}=\widehat{{\bf 1}}_Ne^{i\phi }.
\end{equation}
This is best seen in the center representation, where we have (\ref{unox}) 
\begin{equation}
{\bf U}_{\cal M}(x)={\bf 1}_N(x)e^{i\phi }=e^{i\phi }f_N(x),  \label{uxident}
\end{equation}
with $f_N(x)$ defined in (\ref{eq:trRT}). For the case where $N$ is an odd
integer, we can perform the quantization on points $X$ that are integer
multiples of $\frac{\tau _x^{\prime }}N$ , as we already discussed$.$ We can
see that if the $B$ matrix has all its elements that are multiples of $\frac
N{\tau _x^{\prime }},$ then the propagator given by (\ref{ugxd})\ has the
form (\ref{uxident}), where $f_N(X)=1$ and $\phi =\varphi _N^{\prime }({\cal M})$.
Hence, the $B$ matrix denotes the identity if all its coefficients are
multiples of $\frac N{\tau _x^{\prime }}.$ This implies through (\ref{mbbeta}%
)\ that the ${\cal M}$ matrix must have the form 
\begin{equation}
{\cal M}={\bf 1\quad }\mbox{mod}(N),
\end{equation}
in agreement with reference \cite{hanay} for the $L=1$ case.

For any matrix ${\cal M}\,$ and for a given odd value of $N$, there is some
smallest integer $k(N)$ such that 
\begin{equation}
{\cal M}^{k(N)}={\bf 1\quad }\mbox{mod}(N)  \label{qpf}
\end{equation}
and in this way 
\begin{equation}
\left[ \widehat{{\bf U}}_{\cal M}^0\right] ^{k(N)}={\bf 1}e^{i\phi (N)}.
\end{equation}
In this sense, we may call the quantum propagator periodic with a period
equal to $k(N)\,$ that we then call the {\it quantum period function} (QPF)
although it is completely determined by the classical map through (\ref{qpf}%
). Note that $k(N)=l_N$, i.e., the quantum and classical period function
coincide for $N$ an odd number, as we would expect. It is now easy to see
that the $N^L\,$ eigenangles $\theta _m$ of the unitary propagator $\widehat{%
{\bf U}}_{\cal M}$ are then restricted to lie on the $k(N)$ possible sites 
\begin{equation}
\left\{ \alpha _j=\left[ \frac{2j\pi +\phi (N)}{k(N)}\right] \right\} ,1\leq
j\leq k(N).  \label{alfa}
\end{equation}

The eigenangle spectrum may be related to the traces of the propagator 
\begin{eqnarray}
{\bf Tr}\left[ \left( \widehat{{\bf U}}_{\cal M}\right) ^l\right]
&=&\sum_{m=1}^{N^L}e^{il\theta _m} \\
&=&\sum_{j=1}^{k(N)}d_je^{il\alpha _j}
\end{eqnarray}
where $d_j$ is the degeneracy of the $j$th site defined by (\ref{alfa}). The
density of states,

\begin{equation}
\rho (\theta )=\sum_{i=l}^{N^L}\sum_{l=-\infty }^\infty \delta (\theta
-\theta _i+2\pi l)
\end{equation}
is clearly invariant under $\theta \rightarrow \theta +2\pi .$ Use the
Poisson summation formula, leads to the trace formula, 
\begin{equation}
\rho (\theta )=\frac{N^L}{2\pi }+\frac 1\pi \mbox{Re}\sum_{l=1}^\infty {\bf %
Tr}\left[ \left( \widehat{{\bf U}}_{\cal M}\right) ^l\right] e^{-il\theta },
\label{romapa}
\end{equation}
which holds for all maps on the $2L$-torus. Now, for the cat map, the fact
that the propagator has periodicity $k(N)$ means that the density of states
can also be written in the form \cite{keat2} 
\begin{equation}
\rho (\theta )=\sum_{l=1}^{k(N)}{\bf Tr}\left[ \left( \widehat{{\bf U}}%
_{\cal M}\right) ^l\right] e^{-il\theta }\sum_{j=-\infty }^\infty \delta (\phi
(N)+2\pi j-k(N)\theta ).  \label{rogato}
\end{equation}
That is, the eigenangles are restricted to the sites $\alpha _j$ in (\ref
{alfa}), with the degeneracy at the $j$th site being given by

\begin{equation}
d_j=\frac 1{k(N)}\sum_{l=1}^{k(N)}{\bf Tr}\left[ \left( \widehat{{\bf U}}%
_{\cal M}\right) ^l\right] e^{-il\alpha _j}.
\end{equation}
This last equation then leads to simple expressions for several important
properties of the eigenangle distribution; for example 
\begin{equation}
\sum_{j=1}^{k(N)}d_j^2=\frac 1{k(N)}\sum_{l=1}^{k(N)}\left| {\bf Tr}\left[
\left( \widehat{{\bf U}}_{\cal M}\right) ^l\right] \right| ^2.
\end{equation}

It is important to note that the $\delta $-functions appear explicitly in (%
\ref{rogato}), because the propagator is periodic. The different powers of
the quantum map in (\ref{rogato}) contribute only to the finite
degeneracies, not to the $\delta $-function form of the trace formula.
Moreover, for all finite $N$, only a finite number of these powers are
required to determine the degeneracy at every available site, and hence give
the whole spectrum. However, $k(N)\rightarrow \infty $ as $N\rightarrow
\infty $, thus more terms are required to determine the spectrum as the
semiclassical limit is approached.

We now take the trace of (\ref{ugcorl}), using (\ref{TRA}), to obtain 
\begin{eqnarray}
{\bf Tr}\left[ \left( \widehat{{\bf U}}_{\cal M}\right) ^l\right] &=&{\bf U}%
_{{\cal M}^l}(\xi =0) \\
&=&2^L\left( \tau ({\cal M}^l)\right) ^{-\frac 32}\sum_{{\bf m\in \Diamond }_\xi
^l}\exp \left[ -i2\pi N\frac 14{\bf m}(\beta ^{(l)}-\widetilde{{\frak J}})%
{\bf m}\right] =\frac{e^{i\varphi _N({\cal M}^l)}}{\sqrt{N}^L},
\end{eqnarray}
where we recognize the action of the periodic orbits in the summation
exponent, so 
\begin{eqnarray}
{\bf Tr}\left[ \left( \widehat{{\bf U}}_{\cal M}\right) ^l\right] &=&2^L\left| \det
({\cal M}^l-1)\right| ^{-\frac 32}\sum_{{\bf m\in \Diamond }_\xi ^l}e^{-i2\pi
N\left[ S_f^l({\bf m})\right] }  \label{guttrf} \\
&=&\frac{e^{i\varphi _N({\cal M}^l)}}{\sqrt{N}^L}.  \label{inter}
\end{eqnarray}

Note that ${\bf m\in \Diamond }_\xi ^l$ implies that we are summing over the
periodic orbits of period $l$. Equation (\ref{guttrf}) represents the {\it %
Gutzwiller-Tabor trace formula} \cite{tabor} for the cat map, though we must
note that, instead of being a semiclassical approximation, it is exact in
this case. We must also note that the expression (\ref{inter}) for the
traces of the propagator implies that in the expression (\ref{rogato}) for
the density of states, we have interference of the different phase factors $%
e^{i\varphi _N({\cal M}^l)}$ for the different powers $l$ of the map.

\subsection*{Examples :}

We will now study some quantum features of the classical examples studied in
section 3. All of them fulfill condition (\ref{Mquanti1}), so that their
symmetric $\beta $ matrix has integer entries. Then quantization can be
performed for all odd values of $N,$ for which the quantum propagator will
have the form 
\begin{equation}
{\bf U}_{\cal M}(\Xi )=\sqrt{\frac 1{N^L}}e^{-i2\pi N\left[ \frac 14\Xi \beta \Xi
\right] }
\end{equation}
in the chord representation. In this case, the chords $\Xi $, defined in (%
\ref{corN}), form a lattice of spacing $\frac 2N.$ In the following, we will
study the quantum period function for these maps.

We take cat maps of two degrees of freedom already studied in section 2.
These include all the possible classical behaviors and we will study their
effect on the QPF defined in (\ref{qpf}).

\subsubsection*{ The Hannay and Berry cat map}

\begin{figure}[th]
\centerline {\epsfxsize=6in \epsfysize=4in \epsffile{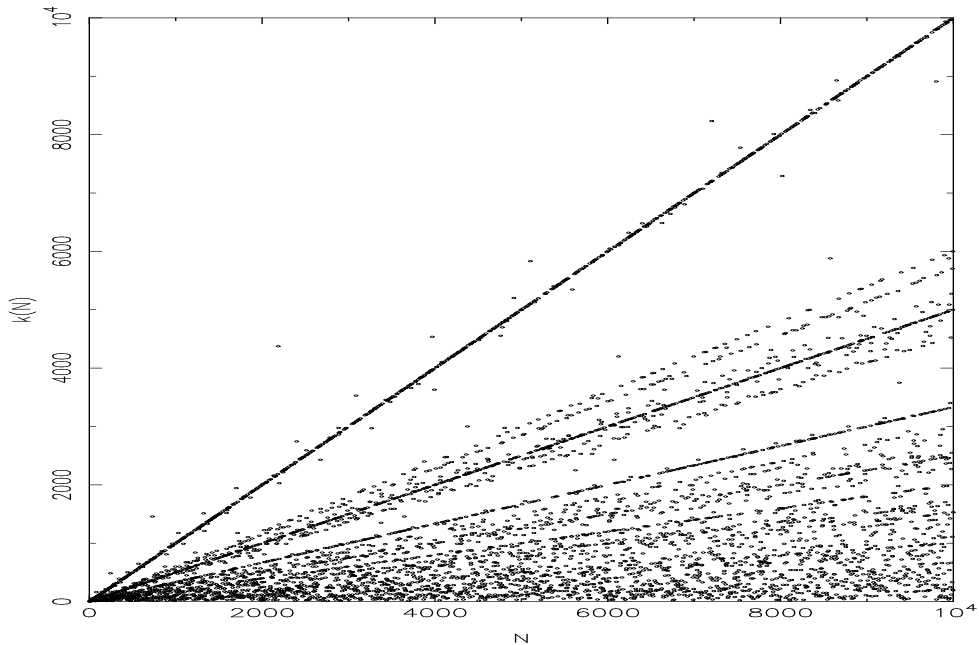} }
\caption{ The QPF for the Hannay and Berry cat map, chaotic with one degree
of freedom, whose symplectic matrix is ${\cal M}_{hb},$ defined in (\ref{mhb}). }
\label{fig.2}
\end{figure}

To compare the different types of behavior that occur for two degrees of
freedom, it is important to present the already known QPF for the Hannay and
Berry cat map whose symplectic matrix is ${\cal M}_{hb}$ defined in (\ref{mhb}).
The QPF\ is shown in figure ~\ref{fig.2}, where we can then see that,
although it has an irregular behavior as a whole, most of the points lie in
families of straight lines which admit a maximum slope. Indeed, this
indicates that there exists an increasing sequence of primes $p$ such that $%
\frac p{k(p)}<C$ for some $C$ independent of $N.$ According to Degli Esposti
et al \cite{degli}, this implies that the quantum map is ergodic and mixing
in the semiclassical limit, within the definition of Von Neumann \cite
{vonneum}. The average behavior of the QPF was established by Keating \cite
{keat2} and it is shown that in the semiclassical limit ($N\rightarrow
\infty $) 
\begin{equation}
\left\langle \log \frac{k(N)}N\right\rangle \approx \frac{-2\sqrt{\pi }}{e^2}%
\left( \log \log N\right) \left( \log \log \log N\right) 
\end{equation}
that is, the degeneracy grows very slowly with $N.$ 

\subsubsection*{ The double hyperbolic case}

\begin{figure}[th]
\centerline {\epsfxsize=6in \epsfysize=4in \epsffile{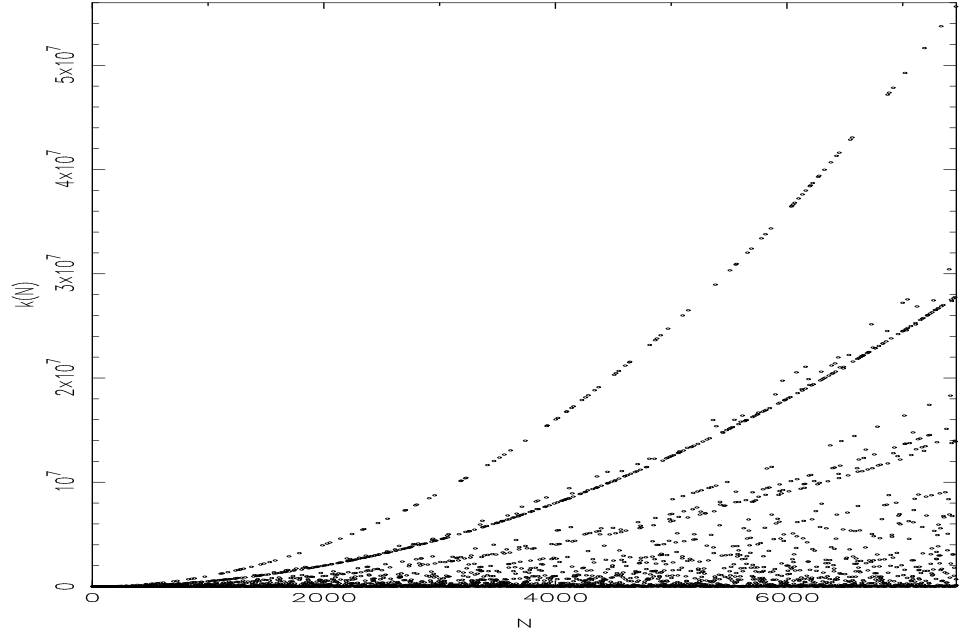} }
\caption{ The QPF for the double hyperbolic map, whose symplectic matrix is $%
{\cal M}_{hh},$ defined in ($\ref{mhh}$). }
\label{fig.3}
\end{figure}

We now choose the symmetric matrix $\beta _{hh}\,$ whose associated
symplectic matrix is given by ${\cal M}_{hh}$ defined in (\ref{mhh}). The QPF of
this map is shown in figure ~\ref{fig.3}. As we can see, the behavior is very
different of that obtained for the Hannay and Berry Cat map. There are now
families of parabolas instead of straight lines. The role played by $N$ in
figure ~\ref{fig.2} is here played by $N^2$, because there are here $N^2$
states for this system. We conjecture that this kind of behavior implies
quantal ergodicity and mixing for two degrees of freedom systems in the
semiclassical limit, although a more formal study of this conjecture will be
realized in a future work. 

\subsubsection*{ The mixed case}

\begin{figure}[tbp]
\centerline {\epsfxsize=6in \epsfysize=3.5in \epsffile{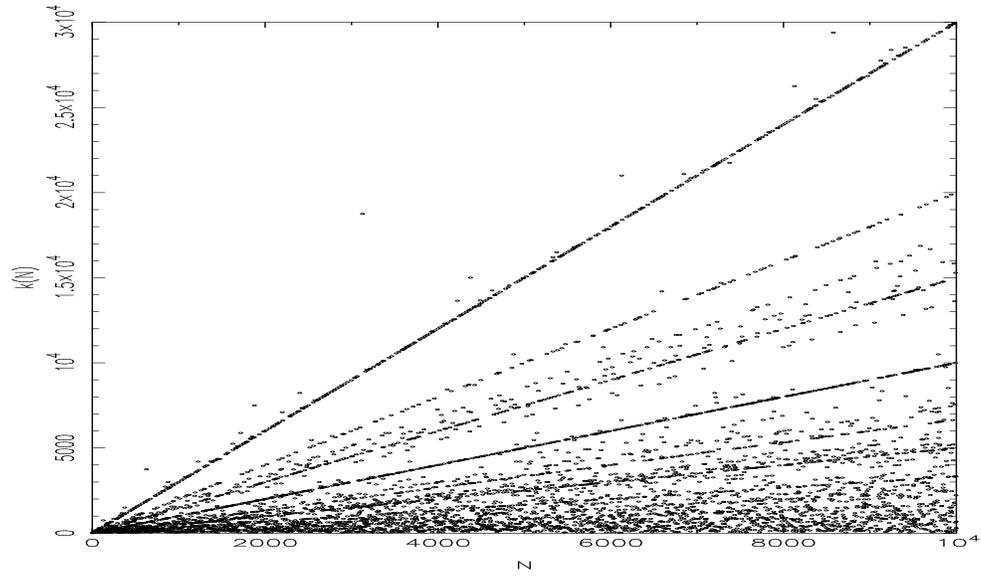} }
\caption{ The QPF for the mixed map , with rational rotations, whose
symplectic matrix is ${\cal M}_{eh1},$ defined in (\ref{meh1}). }
\label{fig.4}
\end{figure}

\begin{figure}[tbp]
\centerline {\epsfxsize=6in \epsfysize=3.5in \epsffile{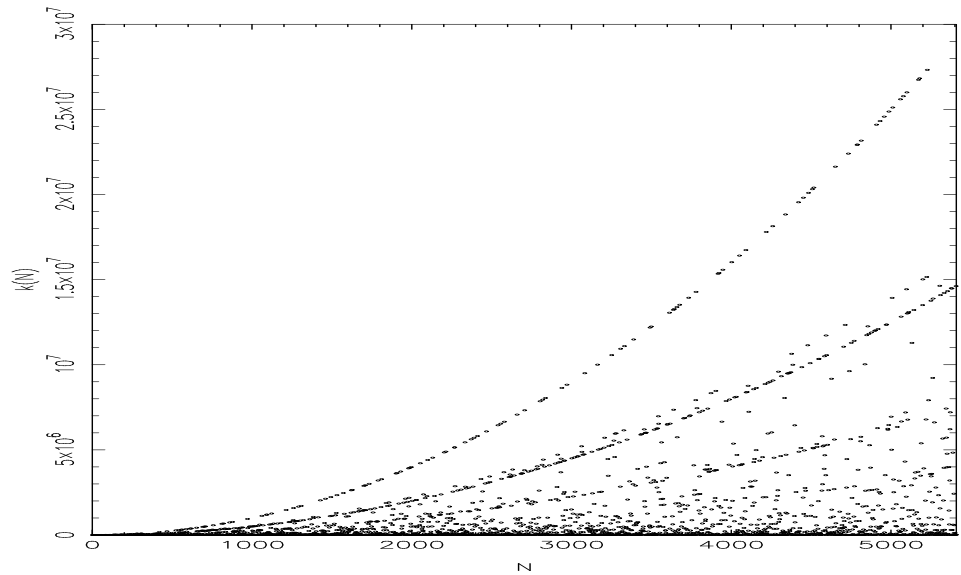} }
\caption{ The QPF for the mixed map, with irrational rotations, whose
symplectic matrix is ${\cal M}_{eh2},$ defined in (\ref{meh2}) .}
\label{fig.5}
\end{figure}

We first consider the map whose symmetric matrix is $\beta _{eh1},$ defined
in (\ref{meh1}), whose QPF is shown in figure ~\ref{fig.4}. This is compared
to $\beta _{eh2}$, defined in (\ref{meh2}), characterized by irrational
rotation angles, whose QPF is now shown in figure ~\ref{fig.5}. There is a
marked difference in the behavior between figure ~\ref{fig.4}. and figure ~\ref
{fig.5}. While the former is very similar to the one obtained in figure ~\ref
{fig.2} for a chaotic one degree of freedom system, the one of figure ~\ref
{fig.5}. is close to figure ~\ref{fig.3}. for a system of two degrees of
freedom. The very different behavior, as was explained in section 3, is due
to the fact that the classical eigenvalues $\lambda _3^{eh1}$ and $\lambda
_4^{eh1}$ denote rotation by angles that are a fraction of $\pi .$ Then the
behavior of the system will be equivalent a hyperbolic system with a single
degree of freedom and it is ergodic only in a subspace. On the contrary, $%
\lambda _3^{eh2}$ and $\lambda _4^{eh2}$ denote rotation angles that are
irrational fractions of $\pi ,$ so that classical and semiclassically the
ergodicity appears in the whole phase space. 
\clearpage

\subsubsection*{ The loxodromic case}

\begin{figure}[h]
\centerline {\epsfxsize=6in \epsfysize=3.5in \epsffile{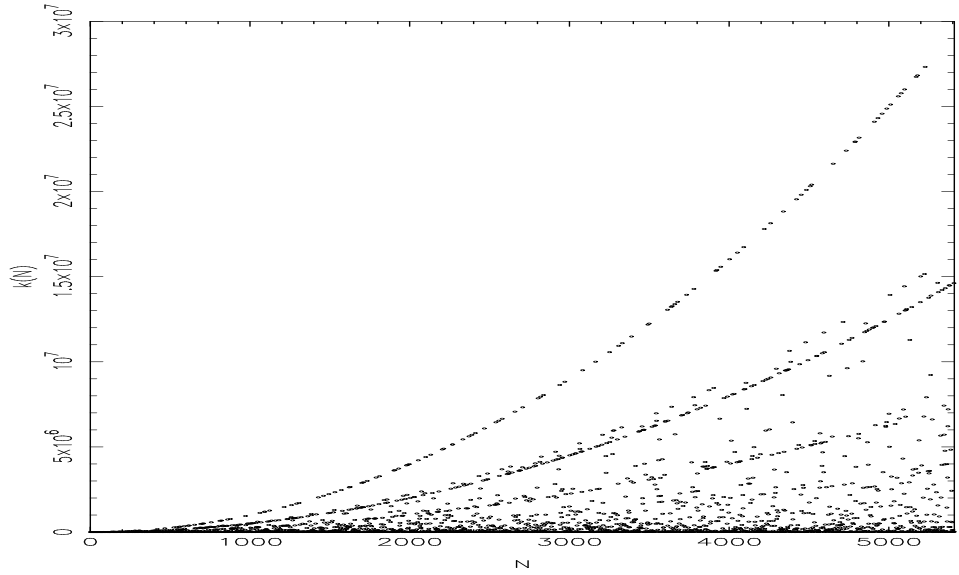} }
\caption{ The QPF for the double loxodromic map of two degrees of freedom,
with irrational rotations, whose symplectic matrix is ${\cal M}_{lox},$ defined in (%
\ref{mlox}).}
\label{fig.6}
\end{figure}

Let us now study the map denoted by the $\beta _{lox}$ matrix defined by (%
\ref{mlox}). The corresponding QPF is shown in figure ~\ref{fig.6}. The
behavior is then similar to the one obtained for ergodic systems with two
degrees of freedom. This map then manifests classical and semiclassical
ergodicity on the whole phase space or the corresponding Hilbert space.
Similarly to the previous example, we found that rotation angles that are a
fraction of $\pi $ behave similarly to hyperbolic one degree of freedom maps
shown in figure ~\ref{fig.5}. 

\clearpage

\section{The quantum feline group}

In this section we present the general quantization of multidimensional cat
maps constructed in section two. In a first step we establish the conditions
for the maps to be quantizable and then we construct their center and chord
representations, based on the formalism developed in reference \cite{opetor}
and summarized in the Appendix.

The classical automorphism generated by ${\cal M}\,$ in plane phase space being
linear, its quantization on the Hilbert space $\left[ {\cal {H}_{{\Bbb R}}}%
\right] ^L$ ,associated to the Euclidean phase space, will have the crucial
property that 
\begin{equation}
\widehat{U}_{\cal M}\widehat{T}_\xi \widehat{U}_{\cal M}^{\dagger }=\widehat{T}_{{\cal M}\xi }
\label{utu}
\end{equation}
for any translation operator $\widehat{T}_\xi $. We now show how to
associate to any $\widehat{U}_{\cal M}$ a unitary operator on $\left[ {\cal H}%
_N^\chi \right] ^L$, the torus Hilbert space characterized by the Floquet
parameter $\chi .$ For this purpose let us study the restriction of $%
\widehat{U}_{\cal M}$ to $\left[ {\cal H}_N^\chi \right] ^L.$ We then have:

{\bf Proposition} 
\begin{equation}
\widehat{U}_{\cal M}\left[ {\cal H}_N^\chi \right] ^L\subset \left[ {\cal H}%
_N^{\chi ^{\prime }}\right] ^L
\end{equation}

{\it where } 
\begin{equation}
\chi ^{\prime }={\cal M}\chi -\frac N2{\frak J}\left( {\cal M}\otimes {\cal M}\right) \quad %
\mbox{mod(1)}  \label{xixip}
\end{equation}

{\it where we have defined the vector product of two matrices as the vector
whose components are } 
\begin{equation}
\left( A\otimes B\right) _i=\sum_{j=1}^LA_{i,j}B_{i,j+L}
\end{equation}

{\bf Proof: }Equation (\ref{utu}) implies 
\begin{equation}
\widehat{U}_{\cal M}\widehat{T}_\xi =\widehat{T}_{{\cal M}\xi }\widehat{U}_{\cal M}.  \label{tuut}
\end{equation}
We now restrict considerations to chords that perform integer loops around
the torus, $\xi ={\bf m,}$ with ${\bf m\,\,}$ an integer vector with $2L$
components,i.e., chords that generate translations that are equivalent to
the identity on the torus. We thus act on a given state of $\left[ {\cal H}%
_N^\chi \right] ^L$ on each side of (\ref{tuut}), so that inserting (\ref
{tNQ}), we obtain 
\begin{equation}
e^{\left[ i2\pi N\left( \frac 14{\bf m}\widetilde{{\frak J}}{\bf m}+{\bf m}%
\wedge \frac \chi N\right) \right] }\widehat{U}_{\cal M}|{\bf \Psi }>=e^{\left[
i2\pi N\left( \frac 14{\bf m}{\cal M}^t\widetilde{{\frak J}}{\cal M}{\bf m}+{\cal M}{\bf m}\wedge 
\frac{\chi ^{\prime }}N\right) \right] }\widehat{U}_{\cal M}|{\bf \Psi }>,
\end{equation}
for any vector ${\bf m}$. Now we chose ${\bf n}={\cal M}{\bf m}$ , hence, we find
that 
\begin{equation}
{\bf n}\wedge \chi ^{\prime }={\cal M}^{-1}{\bf n}\wedge \chi +\frac N4{\bf n}%
\left[ \left( {\cal M}^{-1}\right) ^t\widetilde{{\frak J}}\left( {\cal M}^{-1}\right) -%
\widetilde{{\frak J}}\right] {\bf n}\qquad \mbox{mod}(1)  \label{xixipk}
\end{equation}
and we specify $2L\,$independent integer vectors ${\bf n}\,=e_j%
\,(j=1,...,2L),$ such that each of them denotes one loop around one of the $j
$-irreducible circuit on the torus. In this case $e_j\widetilde{{\frak J}}%
e_j=0,$ using the symplectic property of ${\cal M}$ (\ref{msimple}), and the fact
that ${\frak J}\widetilde{{\frak J}}{\frak J}=\widetilde{{\frak J}}$, we
obtain 
\begin{equation}
e_j\wedge \chi ^{\prime }={\cal M}^{-1}e_j\wedge \chi +\frac N4e_j{\frak J}{\cal M}%
\widetilde{{\frak J}}{\cal M}^t{\frak J}e_j\mbox{mod(1)}.
\end{equation}
Remarking that 
\begin{equation}
\frac N4\left( {\frak J}{\cal M}\widetilde{{\frak J}}{\cal M}^t{\frak J}\right) _i=-\frac
N2\sum_{j=1}^L{\cal M}_{i,j}{\cal M}_{i,j+L},
\end{equation}
we finally obtain (\ref{xixip}) after straightforward manipulations.

In the case $L=1$, (\ref{xixip}) reduces to the already known result \cite
{degli},\cite{debievre}, 
\begin{equation}
\chi ^{\prime }={\cal M}\chi -\frac N2{\frak J}\left( 
\begin{array}{l}
{\cal M}_{11}{\cal M}_{12} \\ 
{\cal M}_{21}{\cal M}_{22}
\end{array}
\right) \quad \mbox{mod(1)}.
\end{equation}
For two degrees of freedom ($L=2$) we obtain:

\begin{equation}
\chi ^{\prime }={\cal M}\chi -\frac N2{\frak J}\left( 
\begin{array}{l}
{\cal M}_{11}{\cal M}_{13}+{\cal M}_{12}{\cal M}_{14} \\ 
{\cal M}_{21}{\cal M}_{23}+{\cal M}_{11}{\cal M}_{24} \\ 
{\cal M}_{31}{\cal M}_{33}+{\cal M}_{32}{\cal M}_{34} \\ 
{\cal M}_{41}{\cal M}_{43}+{\cal M}_{42}{\cal M}_{44}
\end{array}
\right) \quad \mbox{mod(1)}.  \label{xixip4}
\end{equation}

Given ${\cal M},$ there exists for each $N$ a set of solutions of (\ref{xixip})
such that $\chi ^{\prime }=\chi $. This Floquet set of parameters is
determined by 
\begin{equation}
\chi =\frac N2\left( {\cal M}-1\right) ^{-1}{\bf i}_{\cal M}+\left( {\cal M}-1\right) ^{-1}{\bf m,%
}  \label{xiquanti}
\end{equation}
where ${\bf i}_{\cal M}={\frak J}\left( {\cal M}\otimes {\cal M}\right) \mbox{mod(4)}\,$ is an
integer vector. For this set of parameters the propagator $\widehat{U}_{\cal M}$ is
such that 
\begin{equation}
\left[ \widehat{U}_{\cal M},\widehat{{\bf 1}}_N^\chi \right] =0,  \label{[u,1]}
\end{equation}
where $\widehat{{\bf 1}}_N^\chi $ is the unit operator in $\left[ {\cal H}%
_N^\chi \right] ^L$ defined in (\ref{eq:1toro})$.$ Evidently we cannot
change the Floquet parameters at each iteration of the map, so (\ref
{xiquanti}) defines the {\it quantizability set of the map}. For this set of
parameters, we can then restrict the unitary operators $\widehat{U}_{\cal M}\,$ to
the Hilbert space of the $2L$-torus $\left[ {\cal H}_N^\chi \right] ^L$ and
thus define the corresponding torus propagator as 
\begin{equation}
\widehat{{\bf U}}_{\cal M}^\chi =\widehat{U}_{\cal M}\widehat{{\bf 1}}_N^\chi =\widehat{%
{\bf 1}}_N^\chi \widehat{U}_{\cal M}\widehat{{\bf 1}}_N^\chi .  \label{ugato}
\end{equation}
Operators defined through (\ref{ugato}) inherit the unitary from their plane
counterpart and we will call them {\it quantum cat maps. }

The fact that the quantum symplectic map commutes with projections on the
torus (\ref{eq:abpt}) leads to; 
\begin{equation}
\widehat{{\bf U}}_{{\cal M}^l}^\chi =\left[ \widehat{{\bf U}}_{\cal M}^\chi \right] ^l.
\label{uml}
\end{equation}
The unitarity of $\widehat{{\bf U}}_{\cal M}^\chi $ and (\ref{uml}) implies that 
\begin{equation}
\widehat{{\bf U}}_{\cal M}^\chi =\left[ \widehat{{\bf U}}_{\cal M}^\chi \right]
^{-1}=\left[ \widehat{{\bf U}}_{\cal M}^\chi \right] ^{\dagger }.
\end{equation}
Having two cat maps ${\cal M}_A$ and ${\cal M}_B$ that are quantizable on the same Floquet
parameter $\chi ,$ implies that (\ref{eq:abpt}) 
\begin{equation}
\widehat{{\bf U}}_{{\cal M}_A}^\chi \widehat{{\bf U}}_{{\cal M}_B}^\chi =\widehat{{\bf U}}%
_{{\cal M}_A{\cal M}_B}^\chi ,  \label{ucompoL}
\end{equation}
that is, the quantization of the composition of different cat maps is
equivalent to the composition of the quantization of each map. This shows
that quantum cat maps form a group, the quantum propagator preserving the
classical composition laws. This{\it \ feline quantum group} is at the same
time a projection onto the torus of the metaplectic group and a subgroup of
the classical feline group. It then follows that, for any power $l$, the map 
${\cal M}^l$ is also quantizable.

The quantizability set of the map can be made independent of $N$ if the
inhomogeneous term in (\ref{xixip})\ vanishes. This will be the case for
matrices ${\cal M}$, such that 
\begin{equation}
{\bf m}\left( {\cal M}^t\widetilde{{\frak J}}{\cal M}\pm \widetilde{{\frak J}}\right) {\bf %
m=}0\qquad \mbox{mod}(2)  \label{Mquanti}
\end{equation}
for any integer vector ${\bf m.}$ This condition is equivalent to 
\begin{equation}
\sum_{j=1}^L{\cal M}_{i,j}{\cal M}_{i,j+L}=\mbox{even},  \label{Mxi0}
\end{equation}
so that the allowed Floquet parameters are 
\begin{equation}
\chi =\left( {\cal M}-1\right) ^{-1}{\bf m.}
\end{equation}
In conclusion, the Floquet parameters are then denoted by the fixed points
of the map (\ref{xfix}). The QPS can thus have any of the fixed points of
the map as its origin; without loss of generality, one can take $\chi =0.$
For the one degree of freedom case (\ref{Mxi0}) implies that the ${\cal M}$ matrix
must be restricted to the family of maps 
\[
\left( 
\begin{array}{cc}
\mbox{even} & \mbox{odd} \\ 
\mbox{odd} & \mbox{even}
\end{array}
\right) \qquad \mbox{or\qquad }\left( 
\begin{array}{cc}
\mbox{odd} & \mbox{even} \\ 
\mbox{even} & \mbox{odd}
\end{array}
\right) 
\]
selected as quantizable by Hannay and Berry \cite{hanay}.

Let us now provide an explicit construction of quantum cat maps based on the
center and chord representations of operators. We start with the
quantization of the linear automorphism ${\cal M}$ on the Hilbert space $\left[ 
{\cal {H}_{{\Bbb R}}}\right] ^L$ associated to the Euclidean phase space.
The linearity of the ${\cal M}$ map implies in the exactness of the Van Vleck
construction of the propagator \cite{van vleck}. This propagator in the
chord representation has the form \cite{ozrep} 
\begin{equation}
U_{\cal M}(\xi )=\left| \det \left[ 1\pm {\frak J}\frac{\partial ^2S(\xi )}{%
\partial \xi ^2}\right] \right| ^{\frac 12}e^{-\frac i\hbar S(\xi )}
\label{upcor}
\end{equation}
where $S(\xi )\equiv S(\xi ,0)$ is the chord generating function of the
automorphism in the plane phase space. In the center representation, the Van
Vleck propagator is 
\begin{equation}
U_{\cal M}(x)=\left| \det \left[ 1\pm {\frak J}\frac{\partial ^2S(x)}{\partial x^2}%
\right] \right| ^{\frac 12}e^{\frac i\hbar S(x)},  \label{upx}
\end{equation}
where now $S(x)\equiv S(x,0)$ is the center generating function of the
transformation in the plane. Inserting (\ref{scor}) and (\ref{sx}) for the
respective generating functions, we have \cite{ozrep} 
\begin{equation}
U_{\cal M}(x)=\left| \det \left( 1\pm {\frak J}B\right) \right| ^{-\frac
12}e^{\frac i\hbar xBx}  \label{ugatox}
\end{equation}
and 
\begin{equation}
U_{\cal M}(\xi )=\left| \det \left( 1\pm {\frak J}\beta \right) \right| ^{-\frac
12}e^{-\frac i\hbar \frac 14\xi \beta \xi }.  \label{ugatocor}
\end{equation}

We may now project these representations on the $2L$-torus. For this purpose
we must restrict our construction to the quantizability set of Floquet
parameter $\chi $ defined through (\ref{xiquanti}) and take the chord and
center symbols of the projected operator. This is performed substituting (%
\ref{ugatocor}) and (\ref{ugatox}) respectively in (\ref{eq:Acorprom}) and (%
\ref{eq:Axplator}). Then we obtain the chord representation of the quantum
cat map as an average over winding vectors: 
\begin{equation}
{\bf U}_{\cal M}^\chi (\xi )=\left| \det \left( 1\pm {\frak J}\beta \right) \right|
^{-\frac 12}\left\langle e^{-i2\pi N\left[ \frac 14\xi \beta \xi +\frac
12\xi (\beta +{\frak J}){\bf m}+\frac 14{\bf m}(\beta -\widetilde{{\frak J}})%
{\bf m+}\frac \chi N\wedge {\bf m}\right] }\right\rangle _{{\bf m}}.
\label{ugatocor2}
\end{equation}
There are various classical objects present in this formula. Recalling
section 2, $S(\xi ,{\bf m}),$ the chord generating function of the cat map
defined in (\ref{scor}) appears in the exponent of (\ref{ugatocor2}). In the
amplitude we recognize $\tau _\xi ,$ the number of fixed points of the map
defined in (\ref{d1}), so that, 
\begin{equation}
{\bf U}_{\cal M}^\chi (\xi )=2^L\frac 1{\sqrt{\tau _\xi }}\left\langle e^{-i2\pi
N\left[ S(\xi ,{\bf m}){\bf +}\frac \chi N\wedge {\bf m}\right]
}\right\rangle _{{\bf m}}.  \label{ugatocors}
\end{equation}
Similarly, for the center representation we obtain 
\begin{equation}
{\bf U}_{\cal M}^\chi (x)=\left| \det \left( 1\pm {\frak J}B\right) \right|
^{-\frac 12}\left\langle e^{i2\pi N\left[ xBx+x(B-{\frak J}){\bf m}+\frac 14%
{\bf m}(B+\widetilde{{\frak J}}){\bf m-}\frac \chi N\wedge {\bf m}\right]
}\right\rangle _{{\bf m}},  \label{ugatox2}
\end{equation}
where we recognize $S(x,{\bf m}),$ the center generating function of the cat
map (\ref{sx}), and $\tau _x,$ the number of orbits centered on $x$ defined
by (\ref{gama}), so that 
\begin{equation}
{\bf U}_{\cal M}^\chi (x)=2^L\frac 1{\sqrt{\tau _x}}\left\langle e^{i2\pi N\left[
S(x,{\bf m}){\bf -}\frac \chi N\wedge {\bf m}\right] }\right\rangle _{{\bf m}%
}.  \label{ugatoxs}
\end{equation}

Both (\ref{ugatocors}) and (\ref{ugatoxs}) are representations of quantum
cat maps of general dimension, showing that the quantum propagator is
entirely defined in terms of classical objects, except for the term $\frac
\chi N\wedge {\bf m}$ in the exponent that denotes the quantum features of
the boundary conditions. However, we have to perform the average on ${\bf m}$
to make the representation explicit. Given that the quantizability set (\ref
{xiquanti}) only admits rational values of $\chi $, (\ref{ugatocors}) and (%
\ref{ugatoxs}) are Gaussian sums. The quantizability condition (\ref{Mquanti}%
) for the map ${\cal M}$ is equivalent to the condition that the Gaussian sums do
not vanish for $L=1.$ We do not know of a similar verification for $L>1.$

In the following we will restrict our attention to maps that fulfill the
condition (\ref{Mquanti}), such that the inhomogeneous term in (\ref{xixip})
vanishes. As we have already discussed, we can then choose $\chi =0$ without
loss of generality. The quantum cat maps associated with these Floquet
parameters will be denoted by $\widehat{{\bf U}}_{\cal M}$. All the examples in
section 4 are of this type.

We now use the periodicity properties (\ref{scorper}) of $S(\xi ,{\bf m})$
to show that the exponential function in (\ref{ugatocors}) is periodic.
Indeed, for ${\bf m}^{\prime }$ defined in (\ref{mprim}) and the generating
function $S(\xi ,{\bf m}^{\prime })$ defined in (\ref{scorper}) we have that 
\begin{equation}
e^{-i2\pi N\left[ S(\xi ,{\bf m}^{\prime })\right] }=e^{-i2\pi N\left[ S(\xi
,{\bf m})\right] }e^{-i2\pi N\left[ \xi \wedge {\bf k-}\frac 12{\bf m}\Gamma
_1{\bf k-}\frac 14{\bf k}\Delta _1{\bf k}\right] },
\end{equation}
where $\Gamma _1$ and $\Delta _1$ were respectively defined in (\ref{gama1})
and (\ref{delta1}). We can now see that $e^{-i2\pi N\left[ \xi \wedge {\bf k}%
\right] }=1$ for any integer vector ${\bf k}$ and for maps that fulfill the
condition (\ref{Mquanti}) $e^{-i2\pi N\left[ {\bf -}\frac 12{\bf m}\Gamma _1%
{\bf k-}\frac 14{\bf k}\Delta _1{\bf k}\right] }=1,$ hence 
\begin{equation}
\exp \left[ -i2\pi NS\left( \xi ,{\bf m}+({\cal M}-1){\bf k}\right) \right] =\exp
\left[ -i2\pi NS(\xi ,{\bf m})\right] .
\end{equation}
The average in (\ref{ugatocors}) is then periodic, so that, we sum over one
period and divide by the number of points in the period, to perform such an
average. Hence, we must restrict ${\bf m}$ to the fundamental parallelogram $%
\Diamond _\xi $ defined in (\ref{para1}), where there are exactly $\tau _\xi 
$ points ${\bf m}$ with integer coordinates: 
\begin{eqnarray}
{\bf U}_{\cal M}(\xi ) &=&2^L\left( \tau _\xi \right) ^{-\frac 32}\sum_{{\bf m\in
\Diamond }_\xi }e^{-i2\pi N\left[ S(\xi ,{\bf m})\right] }  \nonumber \\
&=&2^L\left( \tau _\xi \right) ^{-\frac 32}\sum_{{\bf m\in \Diamond }_\xi
}e^{-i2\pi N\left[ \frac 14\xi \beta \xi +\frac 12\xi (\beta +{\frak J}){\bf %
m}+\frac 14{\bf m}(\beta -\widetilde{{\frak J}}){\bf m}\right] }.
\label{ugcorp}
\end{eqnarray}
Since ${\bf m}$ belongs to $\Diamond _\xi ,$ the sum in (\ref{ugcorp}) is
the sum over the different classical orbits whose chord is $\xi $.

In a similar way, for the center representation the periodicity region is
the parallelogram $\Diamond _x\,,$ defined in (\ref{para2}), that has
exactly $\tau _x$ points ${\bf m}$ with integer coordinates (\ref{gama}), so
that 
\begin{eqnarray}
{\bf U}_{\cal M}(x) &=&2^L\tau _x^{-\frac 32}\sum_{{\bf m\in \Diamond }_x}e^{i2\pi
N\left[ S(x,{\bf m})\right] }  \nonumber \\
&=&2^L\tau _x^{-\frac 32}\sum_{{\bf m\in \Diamond }_x}e^{i2\pi N\left[
xBx+x(B-{\frak J}){\bf m}+\frac 14{\bf m}(B+\widetilde{{\frak J}}){\bf m}%
\right] },  \label{ugxp}
\end{eqnarray}
i.e. we are taking the sum on the different classical orbits centered in $x.$
The expressions (\ref{ugcorp}) and (\ref{ugxp}) have exactly the form
expected for the semiclassical approximation of the propagator in center and
chord representations respectively \cite{ozrep},\cite{opetor} but in this
case they are exact instead of being a mere approximation.

For generic cat maps, (\ref{ugcorp}) and (\ref{ugxp}) generate Gaussian
sums, but in some cases important simplifications are possible. Let us
start, once more, with the chord representation. We have already seen that
the matrix $\beta $ has the form $\beta =\frac{\overline{\beta }^{\prime }}{%
\tau _\xi ^{\prime }}$, where the barred matrix has integer elements$.$ The
simplest case is when the cat map is such that the associated $\beta \,$%
matrix itself has integer entries. Recalling that $\xi =\frac 1N\overline{%
\xi }$, we will transform to an equivalent set of chords, for which the
values for $\overline{\xi }$ are even multiples of $\tau _\xi ^{\prime }$.
So, for any chord $\xi $ there is an equivalent chord $\Xi ,$ such that 
\begin{equation}
\Xi =\xi +{\bf n}=\frac{2\tau _\xi ^{\prime }}N\overline{\Xi },  \label{corN}
\end{equation}
where the components of $\overline{\Xi }$ are integer numbers up to $N$ and
the components of ${\bf n}$ are integer up to $2\tau _\xi ^{\prime }-1$.
Equation (\ref{corN}) has solutions with the specified features for any $\xi
,$ only if $N\,$ and $2\tau _\xi ^{\prime }\,$ are coprime numbers. We will
then restrict $N$ to be an odd integer. In this case, the chords $\Xi $ in (%
\ref{corN}) form a lattice with spacing $\frac{2\tau _\xi ^{\prime }}N.$ A
hypercube of side $2\tau _\xi ^{\prime }$ has then $N^L\times N^L$
successive chords $\Xi $ that constitute a basis for translation operators.
For the simplest case where the matrix $\beta $ has integer entries, the
chords $\Xi \,$ form a lattice of length $2$ with spacing $\frac 2N.$
Performing the transformation to chords $\Xi ,$\ we see that the term $%
e^{-i2\pi N\left[ \frac 12\Xi (\beta +{\frak J}){\bf m}\right] }=1,$ so
there is then no $\Xi $-dependence in the propagator sum (\ref{ugcorp}),
leading to(\ref{ugcorimp}).

In the same way, the matrix $B$ in the center representation has the form $B=%
\frac{\overline{B}^{\prime }}{\tau _x^{\prime }}$ . Then we transform to
center points $X$ that are integer multiples of $\frac{\tau _x^{\prime }}N$, 
\begin{equation}
X=x+\frac 12{\bf j}=\frac{\tau _x^{\prime }}N\overline{X},  \label{xN}
\end{equation}
where the components of $\overline{X}$ are integer numbers up to $N$ and the
components of ${\bf j}$ are integer up to $2\tau _x^{\prime }-1$. Again,
solutions of (\ref{xN}) with the specified features will exist only if $N$
and $2\tau _x^{\prime }$ are coprime numbers. Thus, reflection operators on
points $X$, that form a lattice with separation $\frac{\tau _x^{\prime }}N$
on a hypercube of side $\tau _x^{\prime },$ form a basis for the Hilbert
space of the torus. For center points $X$ in (\ref{xN}) , the term $e^{i2\pi
N\left[ X(B-{\frak J}){\bf m}\right] }=1,$ so the propagator (\ref{ugxp})
has the form (\ref{ugxd}).

The above are then special cases where the propagator on the torus has the
same form as its equivalent on the plane. These cases are then ideal to
study quantization, as we have seen in section 4.

We will now discuss the quantum effect of a similarity transformation of the
form 
\begin{equation}
{\cal M}\rightarrow {\cal M}^{\prime }={\cal N}^{-1}{\cal M}{\cal N}.
\end{equation}
As we have already discussed the matrices ${\cal M}^{\prime }$ and ${\cal M}$ represent
the same map, but seen on a different frame of canonical coordinates. On the
torus, we have to restrict both ${\cal M}$ and ${\cal N}$ to have integer elements so that
the torus is mapped on itself. This restricts the symplectic similarity
transformations to the feline transformations.

The main advantage of the chord and center representation in plane phase
space is their symplectic invariance\cite{ozrep}. It is well known that
linear classical canonical transformations $x^{\prime }={\cal N}x$ correspond to
unitary transformations in $\left[ {\cal {H}_{{\Bbb R}}}\right] ^L$, 
\begin{equation}
\widehat{A}\rightarrow \widehat{A}^{\prime }=\widehat{U}_{\cal N}\widehat{A}%
\widehat{U}_{\cal N}^{-1}.
\end{equation}
The effect of such a unitary transformation on the chord and center
representation is merely 
\begin{equation}
A(x)\rightarrow A({\cal N}x)\quad \mbox{and}\quad A(\xi )\rightarrow A({\cal N}\xi ).
\end{equation}

Because of the commutation of operator products with projection from the
plane to the torus, the effect of a similarity transformation $\widehat{{\bf %
A}}\rightarrow \widehat{{\bf U}}_{\cal N}^\chi \widehat{{\bf A}}\left[ \widehat{%
{\bf U}}_{\cal N}^\chi \right] ^{-1}$ performed by a quantized cat map on any
operator $\widehat{{\bf A}}$ that commutes with $\widehat{{\bf 1}}_{\cal N}^\chi $
will be purely classical in the center or the chord representations: 
\begin{equation}
{\bf A}(x)\rightarrow {\bf A}({\cal N}x)\quad \mbox{and\quad }{\bf A}(\xi
)\rightarrow {\bf A}({\cal N}\xi ).  \label{cat1}
\end{equation}
Thus the similarity transformation among quantum cat maps, reduces to the
classical similarity transformation: 
\begin{equation}
\widehat{{\bf U}}_{{\cal M}^{\prime }}^\chi =\widehat{{\bf U}}_{\cal N}^\chi \widehat{{\bf %
U}}_{\cal M}^\chi \left[ \widehat{{\bf U}}_{\cal N}^\chi \right] ^{-1},  \label{uprim}
\end{equation}
so that, 
\begin{equation}
{\bf U}_{{\cal M}^{\prime }}(x)={\bf U}({\cal N}x)\quad \mbox{and\quad }{\bf U}_{{\cal M}^{\prime
}}(\xi )={\bf U}({\cal N}\xi ).  \label{catcat}
\end{equation}
However, the quantum cat maps ${\cal M}^{\prime }$ and ${\cal M}$ must be quantized on the
same Floquet parameters $\chi $. This imposes another restriction on the
matrix ${\cal N}$ used to perform the similarity transformation, its quantizability
set must include the Floquet parameter $\chi .$ That is ${\cal M}$ and ${\cal N}$ must
belong the same quantum feline group.

For two quantum cat maps $\widehat{{\bf U}}_{{\cal M}^{\prime }}^\chi $ and $%
\widehat{{\bf U}}_{\cal M}^\chi $ related by (\ref{uprim}) and for all power $l$ of
the map we have 
\begin{eqnarray}
Tr\left[ \left( \widehat{{\bf U}}_{{\cal M}^{\prime }}^\chi \right) ^l\right]
&=&Tr\left\{ \left( \widehat{{\bf U}}_{\cal N}^\chi \widehat{{\bf U}}_{\cal M}^\chi \left[ 
\widehat{{\bf U}}_{\cal N}^\chi \right] ^{-1}\right) ^l\right\} =Tr\left\{ \left( 
\widehat{{\bf U}}_{\cal M}^\chi \right) ^l\left[ \widehat{{\bf U}}_{\cal N}^\chi \right]
^{-l}\left( \widehat{{\bf U}}_{\cal N}^\chi \right) ^l\right\}  \nonumber \\
&=&Tr\left[ \left( \widehat{{\bf U}}_{\cal M}^\chi \right) ^l\right] .  \label{TRML}
\end{eqnarray}
We have seen in (\ref{romapa}) that the density of states and hence the
spectrum of the system is uniquely determined through the traces of the
different powers of the map. Then, (\ref{TRML}) shows that $\widehat{{\bf U}}%
_{{\cal M}^{\prime }}^\chi $ and $\widehat{{\bf U}}_{\cal M}^\chi $ related by (\ref{uprim}%
) have the same quasi-energy spectrum.

We now show that the quantum period function is also invariant with respect
to a feline similarity transformation, for the periodic case where we can
chose the Floquet parameter $\chi =0$. Suppose that the QPF corresponding to 
$\widehat{{\bf U}}_{{\cal M}^{\prime }}^\chi $ and $\widehat{{\bf U}}_{\cal M}^\chi $ are
respectively $k_{{\cal M}^{\prime }}(N)$ and $k_{\cal M}(N)$ , then 
\begin{equation}
\left( {\cal M}^{\prime }\right) ^{k_{\cal M}(N)}=\left( {\cal N}^{-1}{\cal M}{\cal N}\right)
^{k_{\cal M}(N)}={\cal N}^{-1}\left( {\cal M}\right) ^{k_{\cal M}(N)}{\cal N}.
\end{equation}
For all points $x$ belonging to the QPS, i.e., a lattice of spacing $\frac 1N
$ on the $2L$-torus $\Box ,$ we have 
\begin{equation}
\left( {\cal M}^{\prime }\right) ^{k_{\cal M}(N)}x={\cal N}^{-1}\left( {\cal M}\right) ^{k_{\cal M}(N)}{\cal N}x=x,
\end{equation}
hence 
\begin{equation}
\left( {\cal M}^{\prime }\right) ^{k_{\cal M}(N)}={\bf 1\quad }\mbox{mod}(N),
\end{equation}
so that 
\begin{equation}
k_{{\cal M}^{\prime }}(N)\leq k_{\cal M}(N).
\end{equation}
In the same way, for all points $x$ belonging to the QPS 
\begin{equation}
\left( {\cal M}\right) ^{k_{{\cal M}^{\prime }}(N)}x={\cal N}\left( {\cal M}^{\prime }\right)
^{k_{{\cal M}^{\prime }}(N)}{\cal N}^{-1}x=x,
\end{equation}
hence 
\begin{equation}
k_{\cal M}(N)\leq k_{{\cal M}^{\prime }}(N).
\end{equation}
Therefore, 
\begin{equation}
k_{{\cal M}^{\prime }}(N)=k_{\cal M}(N),
\end{equation}
$\,$ that is, the QPF of both quantum maps $\widehat{{\bf U}}_{{\cal M}^{\prime
}}^\chi $ and $\widehat{{\bf U}}_{\cal M}^\chi $ coincide.

\section{Conclusions}

\setcounter{equation}{0}

In this work we have studied classical and quantum properties of
multidimensional Cat maps. In a first step the classical study was performed
using the symplectically invariant center and chord generating functions.
They allow us to represent the symplectic matrix by a symmetric one. This is
the basis for a complete classification of generic four dimensional cat
maps. Clearly, the advantage of working with the appropriate generating functions will be even more pronounced for cat maps of higher dimension. Two degrees of freedom are sufficient to obtain all distinct types of
dynamics. Loxodromic behavior appears as a new alternative with respect to
usual cat maps with one degree of freedom.

The quantization of cat maps was performed using the recently developed Weyl
representation and its conjugate chord representation. The semiclassical
approximation is exact whatever the number of degrees of freedom or the
characteristics of the cat map. The spectral properties show the same kind
of ''pathologies '' observed for systems with one degree of freedom. Through
the quantum periodicity function, we have indication of quantal ergodicity
and mixing in the semiclassical limit for systems that present this
classical property. We must note that this is one of the first times that
loxodromic behavior is quantized \cite{pedrao}.

According to Anosov's theorem, all cases of fully ergodic classical maps are
structurally stable, that is, a weak nonlinear perturbation leads to a map
whose orbits are topologically equivalent to the original cat map. The
possibility of quantizing such an Anosov map is in no way restricted to one
degree of freedom. In this way, one can obtain continuous families of
quantum torus maps, corresponding to fully chaotic classical maps for each
type of map ( doubly hyperbolic, loxodromic, etc.). These nonlinear maps
will probably avoid the spectral anomalies due to quantum periodicity, as
was verified for the case of a single degree of freedom \cite{matos} ,\cite{boas}.

{\it Acknowledgments: }We thanks helpful discussions with J.P. Keating, who provide us Greenman preprint,  E. Pujals and G. Contreras who provide us Ma\~ne's work. We acknowledge financial support from Pronex-MCT 
and A.M.F.R. also thanks support from CLAF-CNPq. This work was partially supported by contracts ANPCYT PICT97-01015, 
CONICET PIP98-420 and EC-931005AR.

\appendix 
\setcounter{equation}{0} \renewcommand{\theequation}{A.\arabic{equation}}

\section{The Weyl quantization on the Torus}

In this appendix we develop the mathematical tools needed for the
quantization of cat maps, based on our previous work on Weyl quantization on
the torus \cite{opetor}.

\subsection{The Hilbert space of the torus}

In a first stage it is important to treat the specification of the Hilbert
space of quantum states, or {\it prequantization}, independently from the
dynamics of the system. That is, we treat the quantum kinematics,
corresponding to the geometrical description of phase space at the classical
level. Just to simplify the notation we limit the presentation for one
degree of freedom systems, since the extension for $L\ne 1$ is trivial.

A complete description for prequantization must include Bloch boundary
conditions : 
\begin{eqnarray}
{\bf \Psi }(q+1) &=&e^{2\pi i\chi _p}{\bf \Psi }(q),  \label{eq1} \\
{\bf \tilde{\Psi}}(p+1) &=&e^{-2\pi i\chi _q}{\bf \tilde{\Psi}}(p)
\label{eq:eq2}
\end{eqnarray}
where 
\begin{equation}
{\bf \tilde{\Psi}}(p)=(2\pi \hbar )^{-1/2}\int e^{-ipq/\hbar }{\bf \Psi }%
(q)dq,  \label{eq:91}
\end{equation}
and $2\pi i\chi _p$ and $2\pi i\chi _q$ are fixed arbitrary Floquet angles;
that is, the prequantization depends on the vector $\chi =(\chi _p,\chi _q)$
whose coordinates are in the range $0\le \chi _q,\chi _q<1$. Solutions to (%
\ref{eq1}) and (\ref{eq:eq2}) with the connection (\ref{eq:91}) only exist
if there is an integer $N$, so that \cite{debievre}

\begin{equation}
\hbar =\frac 1{2\pi N}.  \label{eq:nh}
\end{equation}
Then the space of solutions spans a Hilbert space ${\cal H}_N^\chi $ having
the finite dimension $N$. Two bases of ${\cal H}_N^\chi $ appear in analogy
with the position and momentum eigenvectors, here denoted by, $|{\bf q}_n>$
and $|{\bf p}_m>$ with $n,m=0,1,\dots ,N-1$. According to the Bloch type
boundary conditions, we will then define the position state on the torus as
an average over equivalent positions in the plane phase space:
\begin{equation}
|{\bf q}_n>=\left\langle |\frac{n+\chi _q}N+k>e^{2\pi ik\chi
_p}\right\rangle _k,  \label{eq:qQ}
\end{equation}
with the Hermitian structure 
\begin{equation}
<{\bf q}_m|{\bf q}_n>_{N,\chi }=<{\bf q}_m|{\bf q}_n>=\delta _{m,n}^{(N)}e^{%
\frac{2\pi i}N(m-n)\chi _p},  \label{eq:QQsim}
\end{equation}
The identity operator in ${\cal H}_N^\chi $ is obtained through 
\begin{equation}
\widehat{{\bf 1}}_N^\chi =\sum_{m=0}^{N-1}|{\bf q}_m><{\bf q}_m|.
\label{eq:1toro}
\end{equation}
In analogy, the momentum eigenvectors 
\begin{equation}
|{\bf p}_m>=\left\langle |\frac{m+\chi _p}N+k>e^{-2\pi ik\chi
_q}\right\rangle _k,
\end{equation}
normalized such that 
\begin{equation}
<{\bf p}_m|{\bf p}_n>=\delta _{m,n}^{(N)}e^{-\frac{2\pi i}N(m-n)\chi _q}.
\end{equation}
The bases are exchanged with the transformation kernel, 
\begin{equation}
<{\bf p}_m|{\bf q}_n>=N^{-1/2}e^{2\pi i(m+\chi _p)(n+\chi _q)/N}\equiv
(F_N^\chi )_{mn},  \label{eq:PQ}
\end{equation}
forming a unitary matrix (finite Fourier transformation). According to (\ref
{eq:nh}) the classical limit corresponds to $N\rightarrow \infty $ .

We then see that positions and momenta form a discrete web on the torus,
that we will call from now on the quantum phase space QPS in accordance
with \cite{galeti1}. For the case of $L$ degrees of freedom, the Hilbert
space of the $2L$-torus is $\left[ {\cal H}_N^\chi \right] ^L={\cal H}%
_N^\chi \times {\cal H}_N^\chi \times ...\times {\cal H}_N^\chi $ ($L$
times). For each degree of freedom there is a grid structure, as above, so
the dimension of the Hilbert space $\left[ {\cal H}_N^\chi \right] ^L$ is $%
N^L.$ In this case the Floquet parameter $\chi =\left( 
\begin{array}{c}
\chi _p \\ 
\chi _q
\end{array}
\right) $ is a $2L$-dimensional vector.

\subsection{Projector of the plane on the Torus:}

We envisage the existence of two Hilbert spaces, one associated to the
torus, ${\cal H}_N^\chi $, and another, ${\cal {H}_{{\Bbb R}}}$, associated
to the Euclidean phase space, formed by the square-integrable distributions
that we extend to quasiperiodic distributions. It is easy to see that ${\cal %
H}_N^\chi $ is the subspace of ${\cal {H}_{{\Bbb R}}}$ that obeys the
boundary conditions (\ref{eq1}).We then define a projection of ${\cal {H}_{%
{\Bbb R}}}$ on ${\cal H}_N^\chi $ through the operator $\widehat{{\bf 1}}%
_N^\chi $,

\begin{equation}
|{\bf \Psi }>=\widehat{{\bf 1}}_N^\chi |\psi >,
\end{equation}
where 
\begin{equation}
\widehat{{\bf 1}}_N^\chi =\sum_{n=0}^{N-1}|{\bf q}_n><{\bf q}_n|.
\end{equation}
We see that $\widehat{{\bf 1}}_N^\chi $ is Hermitian $\widehat{{\bf 1}}%
_N^\chi =\left[ \widehat{{\bf 1}}_N^\chi \right] ^{\dagger }$ and, inserting
(\ref{eq:QQsim}), we obtain that it is a projector 
\begin{equation}
\widehat{{\bf 1}}_N^\chi \widehat{{\bf 1}}_N^\chi =\widehat{{\bf 1}}_N^\chi .
\end{equation}
For all operators $\widehat{A}$ acting in ${\cal {H}_{{\Bbb R}}}$ there is a
projected torus operator $\widehat{{\bf A}}^\chi $ which acts on ${\cal H}%
_N^\chi $ : 
\begin{equation}
\widehat{{\bf A}}^\chi =\widehat{{\bf 1}}_N^\chi \widehat{A}\widehat{{\bf 1}}%
_N^\chi 
\end{equation}
Let us now suppose that the operator $\widehat{A}$ transforms ${\cal H}%
_N^\chi $ into itself, i.e. 
\begin{equation}
\lbrack \widehat{{\bf 1}}_N^\chi ,\widehat{A}]=0,
\end{equation}
then 
\begin{equation}
\widehat{{\bf A}}^\chi =\widehat{{\bf 1}}_N^\chi \widehat{A}\widehat{{\bf 1}}%
_N^\chi =\widehat{A}\widehat{{\bf 1}}_N^\chi 
\end{equation}
and $\widehat{A}$ is said to be a torus invariant operator.

For two operators of this kind: 
\begin{equation}
\widehat{{\bf A}}^\chi \widehat{{\bf B}}^\chi =\widehat{{\bf 1}}_N^\chi 
\widehat{A}\widehat{{\bf 1}}_N^\chi \widehat{{\bf 1}}_N^\chi \widehat{B}%
\widehat{{\bf 1}}_N^\chi =\widehat{{\bf 1}}_N^\chi \widehat{A}\widehat{{\bf 1%
}}_N^\chi \widehat{B}\widehat{{\bf 1}}_N^\chi =\widehat{{\bf 1}}_N^\chi 
\widehat{A}\widehat{B}\widehat{{\bf 1}}_N^\chi =\widehat{A}\widehat{B}%
\widehat{{\bf 1}}_N^\chi .  \label{eq:abpt}
\end{equation}
So, the product of the projected operators is the same as the projection of
the product of operators. For many degrees of freedom, we will simplify the
notation by using $\widehat{{\bf 1}}_N^\chi \,$ to denote both the projector
on $\left[ {\cal H}_N^\chi \right] ^L\,$and on ${\cal H}_N^\chi .$ No
confusion is made since we will always specify the number $L$ of degrees of
freedom we are working in.

\subsection{Restriction of the translations and reflections to the torus}

Projecting the translation and reflection operators on the torus we obtain 
\cite{opetor} 
\begin{equation}
\widehat{{\bf 1}}_N^\chi \widehat{T}_\xi \widehat{{\bf 1}}_N^\chi =\left\{ 
\begin{tabular}{lll}
$\widehat{{\bf T}}_\xi ^\chi $ & $\mbox{if there are }r$ and $s$ integers so
that & $\xi =\left( \frac rN,\frac sN\right) $ \\ 
$0$ & otherwise & 
\end{tabular}
\right. 
\end{equation}
where the torus translation operators $\widehat{{\bf T}}_\xi ^\chi \equiv 
\widehat{{\bf T}}_{r,s}^\chi $ are defined through 
\begin{equation}
\widehat{{\bf T}}_{r,s}|{\bf q}_n>=e^{i\frac{2\pi }Nr(n+\chi _q+s/2)}|{\bf q}%
_{n+s}>\qquad \widehat{{\bf T}}_{r,s}|{\bf p}_m>=e^{i\frac{2\pi }Ns(m-\chi
_p+r/2)}|{\bf p}_{m+r}>.  \label{eq:TQ}
\end{equation}
Their interpretation as {\it Translation operators in QPS} is clear with
this last expression. The $\chi $ dependence is from now on implicit.

Performing ${\bf m}=\left( 
\begin{array}{l}
m_p \\ 
m_q
\end{array}
\right) $ integer loops around the irreducible circuits of the torus we get 
\begin{equation}
\widehat{{\bf T}}_{{\bf m}}=e^{i2\pi N\left[ \left( \frac \chi N\right)
\wedge {\bf m+}\frac 14{\bf m}\widetilde{{\frak J}}{\bf m}\right] }\widehat{%
{\bf 1}}_N^\chi ,  \label{tNQ}
\end{equation}
so that the symmetries of the translation operator are 
\begin{equation}
\widehat{{\bf T}}_{\xi +{\bf m}}=e^{i2\pi N\left[ \left( \frac \xi 2-\frac
\chi N\right) \wedge {\bf m+}\frac 14{\bf m}\widetilde{{\frak J}}{\bf m}%
\right] }\widehat{{\bf T}}_\xi .  \label{eq:TTsim}
\end{equation}
To obtain a basis of operators, $r$ and $s$ must then run in $[0,N-1]$, that
is, we only need translations that perform less than one loop around the
torus.

Similarly, projecting reflection operators, we have 
\begin{equation}
\widehat{{\bf 1}}_N^\chi \widehat{R}_x\widehat{{\bf 1}}_N^\chi =\left\{ 
\begin{tabular}{lll}
$\widehat{{\bf R}}_{x_{a,b}}$ & $\mbox{if there are }a$ and $b$
semi-integers so that & $x=\left( \frac{a+\chi _p}N,\frac{b+\chi _q}N\right) 
$ \\ 
$0$ & otherwise & 
\end{tabular}
\right.
\end{equation}
where 
\begin{equation}
\widehat{{\bf R}}_{x_{a,b}}=\frac
1{2N}\sum_{r=0}^{2N-1}\sum_{s=0}^{2N-1}e^{-i2\pi Nx\wedge \xi }\widehat{{\bf %
T}}_\xi  \label{eq:RRT}
\end{equation}
are the torus reflection operators on the center point $x_{a,b}.$ The
unitarity of $\widehat{{\bf R}}_x$ is ensured by the action of the operators
on the Hilbert space ${\cal H}_N^\chi $ 
\begin{equation}
\widehat{{\bf R}}_x|{\bf q}_n>=e^{i\frac{2\pi }N2(b-n)(a+\chi _p)}|{\bf q}%
_{2b-n}>\qquad \mbox{ and  }\qquad \widehat{{\bf R}}_x|{\bf p}_m>=e^{i\frac{%
2\pi }N2(a-m)(b+\chi _q)}|{\bf p}_{2a-m}>.  \label{eq:RQ}
\end{equation}
Then $\widehat{{\bf R}}_x$ reflects the QPS web about the point $x=(\frac{%
a+\chi _p}N,\frac{b+\chi _q}N)=(x_p,x_q)$. We then need to include
half-integer values of $a$ and $b$ to perform these reflections. Symmetry
properties for reflection operators on points differing in half loops are, 
\begin{equation}
\widehat{{\bf R}}_{x+\frac{{\bf m}}2}=(-1)^{bm_p+am_q+m_pm_qN}\widehat{{\bf R%
}}_x=e^{i2\pi N\left[ \frac \chi N\wedge {\bf m+}\frac 14{\bf m}\widetilde{%
{\frak J}}{\bf m}\right] }\widehat{{\bf R}}_x  \label{eq:symR}
\end{equation}

The $N^2$ independent operators needed for a basis of ${\cal H}_N^\chi $ are
obtained with $\left( a,b\right) $ belonging to ${0,1/2,\dots ,\frac{N-1}2,}$%
i.e., only one quarter of the torus is needed. This is named the Weyl Phase
Space WPS. The traces of the translation and reflection operators are: 
\begin{equation}
{\bf Tr}(\widehat{{\bf T}}_\xi )=Ne^{i\frac{2\pi }N(\frac{rs}2+r\chi
_p-s\chi _p)}\delta _{r,0}^{(N)}\delta _{s,0}^{(N)}\equiv N\;e^{i\frac{2\pi }%
N(\frac{rs}2+r\chi _q-s\chi _p)}\;\delta _\xi ^{(N)}  \label{eq:trT}
\end{equation}
and 
\begin{eqnarray}
{\bf Tr}(\widehat{{\bf R}}_x) &=&f_N(x)=\frac
12(1+(-1)^{2a}+(-1)^{2b}+(-1)^{2a+2b+N}) \\
&=&\left\{ 
\begin{tabular}{ll}
$0$ & $\mbox{ if N is even and }a\mbox{ or }b\mbox{ half-integers}$ \\ 
$2$ & $\mbox{ if N is even and }a\mbox{ and }b\mbox{ integers}$ \\ 
$1$ & $\mbox{if N is odd and }a\mbox{ or }b\mbox{ integers}$ \\ 
$-1$ & $\mbox{ if N is odd and }a\mbox{ and }b\mbox{ half-integers}$%
\end{tabular}
\right.   \label{eq:trRT}
\end{eqnarray}

\subsection{Operators and their Symbols}

With the set of torus translation and reflection operators, we can represent
any operator in ${\cal H}_N^\chi $ . The chord representation of an operator 
$\widehat{{\bf A}}$ is defined as; 
\begin{equation}
{\bf A}(\xi )={\bf Tr}\left( \widehat{{\bf A}}\widehat{{\bf T}}_{-\xi
}\right) .  \label{eq:Acor}
\end{equation}
From the symbol, we recover the operator 
\begin{equation}
\widehat{{\bf A}}=\frac 1N\sum_{r,s=0}^{N-1}{\bf A}(\xi )\widehat{{\bf T}}%
_\xi \equiv \frac 1N\sum_\xi {\bf A}(\xi )\widehat{{\bf T}}_\xi .
\label{eq:corA}
\end{equation}
Performing ${\bf m}$ loops around the torus, the symbol becomes 
\begin{equation}
{\bf A}(\xi +{\bf m})=e^{i2\pi N\left[ \left( \frac \xi 2-\frac \chi
N\right) \wedge {\bf m+}\frac 14{\bf m}\widetilde{{\frak J}}{\bf m}\right] }%
{\bf A}(\xi ).  \label{eq:Acorsim}
\end{equation}
The center representation on the torus of an operator $\widehat{{\bf A}}$ is
defined as 
\begin{equation}
{\bf A}(x)={\bf Tr}\left( \widehat{{\bf A}}\widehat{{\bf R}}_x\right) .
\label{eq:Acen}
\end{equation}
From the symbol we recover the operator through: 
\begin{equation}
\widehat{{\bf A}}=\frac 1N\sum_{a,b=0}^{\frac{N-1}2}\widehat{{\bf R}}_x{\bf A%
}(x)\equiv \frac 1N\sum_x\widehat{{\bf R}}_x{\bf A}(x).  \label{eq:cenA}
\end{equation}
The symmetry properties of $\widehat{{\bf R}}_x$ (\ref{eq:symR}) implies in
, 
\begin{equation}
{\bf A}(x+\frac{{\bf m}}2)=e^{i2\pi N\left[ \frac \chi N\wedge {\bf m+}\frac
14{\bf m}\widetilde{{\frak J}}{\bf m}\right] }{\bf A}(x).  \label{eq:Acensim}
\end{equation}

From the chord or center representations we obtain the more familiar
position representation of the propagator as 
\begin{equation}
{\bf A(q}_m,{\bf q}_n)=\frac 1N\sum_{\xi _p=0}^{N-1}{\bf A}(\xi _p,\xi
_{m-n})e^{i2\pi N\xi _p(\frac{q_m+q_n}2)},  \label{eq:AQcor}
\end{equation}
or 
\begin{equation}
{\bf A(q}_m,{\bf q}_n)=\frac 1N\sum_{x_p=0}^{\frac{N-1}2}{\bf A}(x_p,x_{%
\frac{m+n}2})e^{i2\pi Nx_p(q_m-q_n)}.  \label{eq:AQx}
\end{equation}
The trace of operators in ${\cal H}_N^\chi $ are obtained as 
\begin{equation}
{\bf Tr}\left( \widehat{{\bf A}}\right) ={\bf A}(\xi =0)=\sum_x{\bf A}%
(x)f_N(x)=\frac 12\sum_{a,b=0}^{N-\frac 12}{\bf A}(x).  \label{TRA}
\end{equation}
The representation of the identity operator on the torus Hilbert space $%
{\cal H}_N^\chi $ has the form 
\begin{equation}
{\bf 1}(x)=f_N(x)\ \ \ \mbox{ and }\ \ \ {\bf 1}(\xi )=N\;\delta _\xi ^{(N)}.
\label{unox}
\end{equation}

For the chord representation to denote a unitary operator $\widehat{{\bf U}}$%
, it has to fulfill the condition

\begin{equation}
\left( \frac 1N\right) \sum_{\xi _1}{\bf U}(\xi _1){\bf U}^{*}(\xi _1-\xi
)e^{-i2\pi N\xi _1\wedge \xi }={\bf 1}(\xi )=N\;\delta _\xi ^{(N)},
\label{unitcor}
\end{equation}
while the center representation of unitary operators is restricted to
symbols such that, 
\begin{equation}
\left( \frac 1N\right) ^2\sum_{x_1,x_2}{\bf U}(x_1){\bf U}^{*}(x_2)e^{i4\pi
N(x-x_1)\wedge (x-x_1)}={\bf 1}(x)=f_N(x).  \label{unitx}
\end{equation}

\subsection{Relation between symbols}

We shall now describe how the symbols on the torus can be obtained from
their counterparts on the plane. The result is directly given for $L$
degrees of freedom. Starting with the chord representation, the torus symbol
at points $\xi =\frac 1N\overline{\xi },$ where $\overline{\xi }$ is a $2L$
integer components vector, is calculated in \cite{opetor}, so that 
\begin{equation}
{\bf A}(\xi )=\left\langle e^{i2\pi N\left[ \left( \frac \xi 2-\frac \chi
N\right) \wedge {\bf m+}\frac 14{\bf m}\widetilde{{\frak J}}{\bf m}\right]
}A\left( \xi +{\bf m}\right) \right\rangle _{{\bf m}}.  \label{eq:Acorprom}
\end{equation}
Note that we have to perform a phase weighted average on equivalent points
to obtain the symbol on the torus. In a similar way the symbols in the
center representation on points $x=\frac 1N\left( \overline{x}+\chi \right) $
are also calculated in \cite{opetor} resulting in 
\begin{equation}
{\bf A}(x)=\left\langle e^{i2\pi N\left[ \frac \chi N\wedge {\bf m+}\frac 14%
{\bf m}\widetilde{{\frak J}}{\bf m}\right] }A\left( x+\frac{{\bf m}}2\right)
\right\rangle _{{\bf m}}.  \label{eq:Axplator}
\end{equation}

\end{document}